\documentclass[%
 reprint,
 amsmath,amssymb,
 aps,
]{revtex4-2}

\usepackage{graphicx}
\usepackage{dcolumn}
\usepackage{bm}

\usepackage[ruled,linesnumbered]{algorithm2e}
\usepackage{ragged2e}
\usepackage{color}
\usepackage{xcolor}
\usepackage{array}
\begin{document}

\preprint{APS/123-QED}

\title{Discover governing differential equations from evolving systems}
\author{Yuanyuan Li}
\affiliation{%
Guangzhou Institute of Technology, Xidian University, Guangzhou 510555, China\\
}%


\author{Kai Wu}
  \email{kwu@xidian.edu.cn}
\affiliation{School of Artificial Intelligence, Xidian University, Xi'an 710071, China \\
}%
\author{Jing Liu}
\affiliation{%
 Guangzhou Institute of Technology, Xidian University, Guangzhou 510555\\ 
}


\date{\today}

\begin{abstract}
Discovering the governing equations of evolving systems from available observations is essential and challenging. In this paper, we consider a new scenario: discovering governing equations from streaming data. 
Current methods struggle to discover governing differential equations with considering measurements as a whole, leading to failure to handle this task.
We propose an online modeling method capable of handling samples one by one sequentially by modeling streaming data instead of processing the entire dataset. The proposed method performs well in discovering ordinary differential equations (ODEs) and partial differential equations (PDEs) from streaming data. Evolving systems are changing over time, which invariably changes with system status. Thus, finding the exact change points is critical. The measurement generated from a changed system is distributed dissimilarly to before; hence, the difference can be identified by the proposed method. Our proposal is competitive in identifying the change points and discovering governing differential equations in three hybrid systems and two switching linear systems.
\end{abstract}

\maketitle


\section{\label{sec:level11}Introduction}
Research on the hidden information in measurements that performs a transformative impact on the discovery of dynamical systems has attracted the attention of scientific researchers \cite{1,2,3,4}. Many canonical dynamics are rooted in conservation laws and phenomenological behaviors in physical engineering and biological science \cite{5}. Enabled by the increasing maturity of sensor technology, data-driven methods have promoted various innovations in describing time series generated from experiments \cite{4}. Due to the complexity of dynamical systems, revealing the underlying governing equations representing the physical laws from time-series data that gives a general description of the spatiotemporal activities is a tremendous challenge \cite{3,li2019discovering,reinbold2019data,maddu2021learning,
somacal2022uncovering,shea2021sindy,xu2021robust,chen2022symbolic}.

In a series of developments, modeling methods for complex systems include empirical dynamic modeling \cite{6}, equation-free modeling \cite{7,8}, and modeling emergent behavior \cite{9}. Discovery patterns contain normal form identification \cite{10}, nonlinear Laplacian spectral analysis \cite{11}, dynamic automatic inference \cite{12}, and nonlinear regression \cite{13}. Overall, popular methods for data-driven discovery of complex systems are mainly based on sparse identification \cite{14,15} and deep neural networks \cite{16,18}, such as sparse identification of nonlinear dynamics (SINDy) \cite{20} and PDE-Net \cite{19}. Existing methods have provided an encouraging performance on systems motivated by unchanging rules or architectures, where novelty and variability are seen as disruptive \cite{21}. 

However, although capable of effectively handling spatiotemporal sequences, the above algorithms are batch learning methodologies whose common problem is ephemeral fitting \cite{6,30}, thereby leading to the abortive modeling for streaming data consecutively generated like water flow. The overwhelming majority of complex systems are constantly running and generating new observations one by one over time (such as financial markets, social networks, and neurological connectivity, etc.) \cite{25,26,27,28}, resulting in failing to update gradients in the SINDy. In real-time systems, we must adapt to different stream parts and produce an immediate output before seeing the next input.

Moreover, several SINDy-based methods have been demonstrated to perform online tasks. Autoencoder SINDy is trained offline and queried online to compute the time evolution of the full order system in different conditions \cite{cai2022online}. \cite{conti2023reduced} claims that the SINDy-based algorithm holds feasibility in online applications due to its few computational time for all cases, owing to a little tuning of parameters and Fast Fourier Transform. However, these methods cannot handle streaming data and discover the change points in evolving systems. The most related work is the method in \cite{messenger2022online}, which processes solution snapshots that arrive sequentially. It is a small batch method. Its performance and training iterations depend on the length of snapshots, which will seriously affect the evaluation of change points. Thus, \cite{messenger2022online} has difficulty finding the change point hidden in solution snapshots on account of the nature of the mini-batch method, which limits the application of this method to switching systems.

To the concrete, we regard the measurements reflecting the system status as invariably changing streaming data. In this article, we present a novel method, called Online-Sparse Identification of Nonlinear Dynamics (O-SINDy), to develop a parsimonious model that most accurately represents the real-time streaming data. Our proposal handles samples sequentially by modeling subsampled spatiotemporal sequences as streaming measurements instead of processing the entire dataset directly. Moreover, O-SINDy can bypass batch storage and intractable cases of combinatorial brute-force search across all possible candidate terms.

Mathematically, we conduct experiments with diverse spatiotemporal measurements generated from dynamical systems to verify the excellent performance of O-SINDy. By recovering extensively representative physical system expressions solely from streaming data, the method is demonstrated to work on various canonical instances, including the nonlinear Lorenz system, Burgers’ equation, reaction-diffusion equation, etc. Experimental results show that our framework provides an online technique for real-time online analysis of complex systems. We suggest that O-SINDy, which overcomes the limitations of batch learning methodologies, is competent for deducing the governing differential equations if sequential measurements of complex dynamic systems are available. 

With the ability to handle streaming data, O-SINDy can ideally cope with the parameter estimation of time-varying nonlinear dynamics and is general enough to detect multiple types of evolutionary patterns. O-SINDy outperforms the state-of-the-art methods in discovering governing differential equations in the evolving two-dimensional damped harmonic oscillators, Susceptible-Infected-Recovered disease model with time-varying transmission rates, and two switching linear systems.

\begin{figure*}[htbp]
\includegraphics[width=0.8\linewidth]{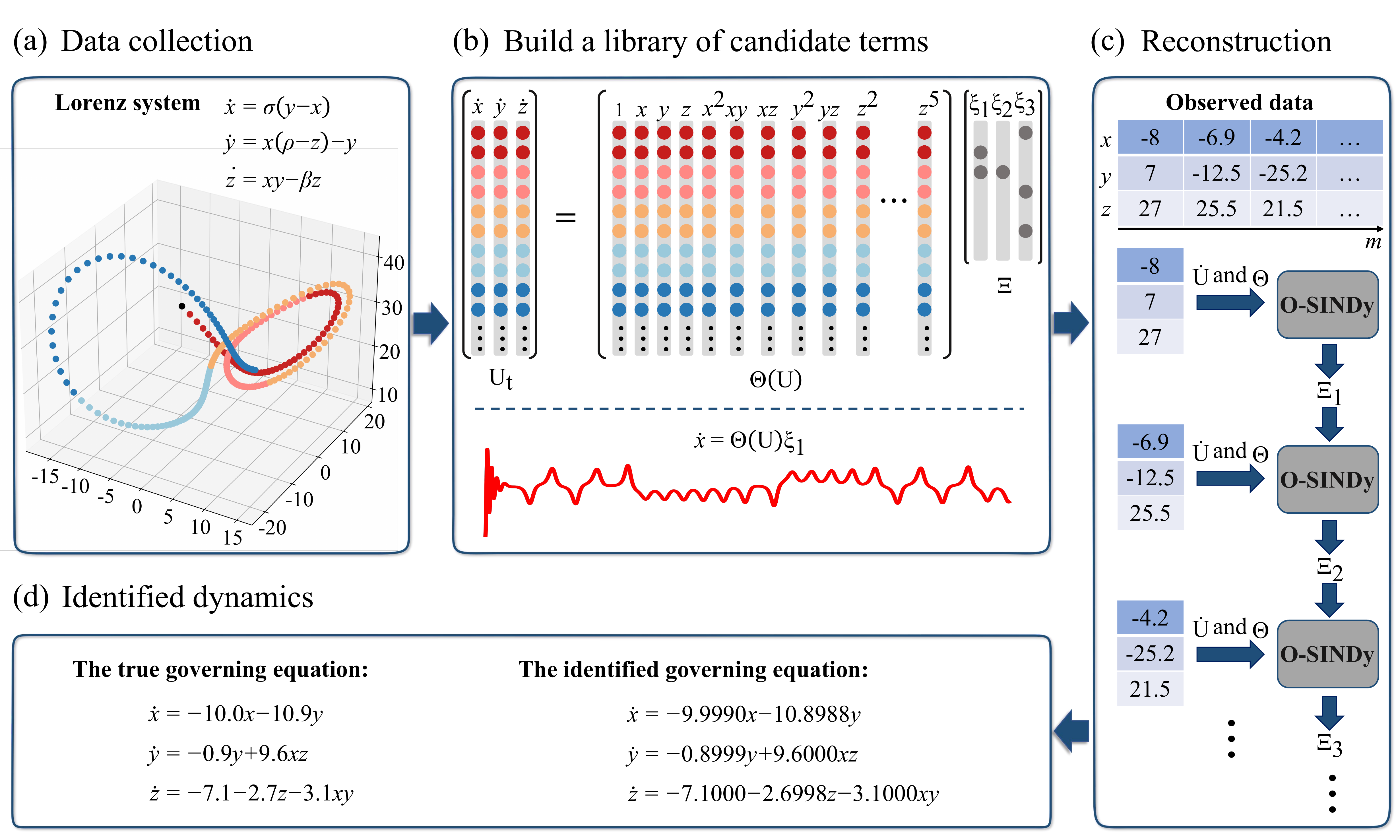}
\caption{Steps in O-SINDy applied to infer the governing equations of the chaotic Lorenz system from streaming data. (a) The particle’s trajectory of motion, starting from the black point. The points of different colors represent the record collected as streams in real-time systems but used as a whole by batch methods. (b) As the observations arrive, take numerical derivatives of the current state vector and construct a library matrix $\Theta$ incorporating the candidate terms for governing equations. (c) Update the form of equations by considering one instance each time. (d) The reconstruction of the Lorenz system.}
\label{fig1}
\end{figure*}
\section{\label{sec:level2}Methods}

With large amounts of data arriving as streams, any offline machine learning algorithm that attempts to store the complete dataset for analysis will fail due to running out of memory. The rise of streaming data raises the technical challenge of a real-time system, which must do any processing or learning encouraged by each record in arriving order. Let $u$ denote the state vector of a real-time system at time $t$, and $u_t$ represent its partial differentiation in time domain. The system receives streaming inputs:
\[\cdots,u_{t-3},u_{t-2},u_{t-1},u_t,u_{t+1},u_{t+2},u_{t+3},\cdots\]
For instance, vector $u$ represents the position information of the particle shown in Fig. 1(a). Every step, the particle moves along the trajectory; the data will be captured by sensors and sent to the system that needs to analyze or respond in time. Instead of storing the entire stream, continuous online learning faces one input $u_t$ at a time.

The rigorous model of a complex system is always a set of differential equations with unknown physics parameters \cite{32}. Without losing generality, we consider a general parameterized physical system of the following form:
\begin{equation}
\label{eq1}
u_t=N\left(1,u,u^2,\ldots,{u_x},uu_x,\ldots,u_{xx},\ldots\right),
\end{equation}
where the terms with subscript $x$ represent partial differentiation of $u$ in space domain, “1” denotes the constant, and $N(\bullet)$ is an unknown combination of the nonlinear functions, partial derivatives, constant, and additional terms of $u(x, t)$. Our goal is to select the correct terms that are most relevant to dynamic information on real-time circumstance. In view of measurements of all considered state variables $U$, the right-hand side of Eq. (\ref{eq1}) can be expressed by multiplying the library function matrix $\Theta\left(U\right)$ with the coefficient matrix $\Xi$ as follows:
\begin{equation}
\label{eq2}
U_t=\Theta\left(U\right)\Xi,
\end{equation}
where
\begin{equation}
\label{eq3}
\Theta\left(U\right)=\left[\theta_1\left(U\right),{\ \theta}_2\left(U\right),\ \ \ldots,\ \theta_p\left(U\right)\right].
\end{equation}

Columns of $\Theta\left(U\right)$, $\theta_s\left(U\right)(s=1,2,\ldots,p)$, correspond to $p$ specific candidate terms for the governing equation, as shown in Fig. \ref{fig1}(b).

O-SINDy is divided into two steps: (i) build complete libraries of candidate terms; (ii) update the structure of governing equations via the FTRL-Proximal style methodology. Each procedure is described as follows.

\subsection{\label{sec:level21}Build a candidate library}
We start by collecting the spatiotemporal series data at $m$ time points and $n$ spatial locations of the state variables. For each state variable, the captured state measurements are represented as a single column state vector $u \in C^{nm \times 1}$. Based on all the observables, e.g., variables $x$, $y$, and $z$ in Lorenz systems, a series of functional terms associated with these quantities can be calculated and then reshaped into a single column as well, such as $x^2$, $xy$, $xyz$, $x^2y$, etc. Likewise, partial differential terms should be considered in the candidate library if PDEs govern the dynamics. Furthermore, “1” denotes the constant term that possibly appears in equations. The compositive function library $\Theta\left(U\right) \in C^{nm \times p}$ is a matrix that contains $p$ designed functional terms. Note that the computed time derivative $u_t \in C^{nm \times 1}$ is also a single-column vector presented on the left-hand side of Eq. (\ref{eq2}). The constant term in the library $\Theta$ sufficiently considers the bias term in the governing differential equation so that the model can be regarded as an unbiased representation of dynamics. For O-SINDy, an important prerequisite for revealing the correct governing differential equation is that the candidate function library contains all the members which constitute the concise dynamics expression, so as to ensure that the exact sparse coefficient $\Xi$ is computed under iterations.

\subsection{\label{sec:level22}Optimization}
	
Formulating a hypothesis that the governing differential equations are evolving at any moment, we model each example as a dynamic process to simulate the arrival of streaming data. Given a set of time-series state measurements at a fixed number of spatial locations in $x$, the goal is to construct the form of $N(\bullet)$ online. To fix this issue, the core of our innovation is utilizing real-time streaming data to denote the loss function. Considering a coefficient vector in $\Xi$, $\xi_j(j=1,2,\ldots,d)$, which corresponds to the specific state variable $u$, the fitness function is designed as follows:
\begin{equation}
\label{eq4}
\min_{\xi_j}=\sum_{i=1}^{mn}L_i(\xi_j)+\lambda_1||\xi_j||_1,
\end{equation}
where $\lambda_1\geq0$ denotes the regularization coefficient, $n$ is the number of positions, $m$ denotes the length of the time series, $\xi_j=\left[\xi_{j1},\xi_{j2},\ldots,\xi_{jp}\right]^T$, and $L_i\left(\xi_j\right)$ is defined as follows:
\begin{equation}
\label{eq5}
L_i\left(\xi_j\right)=\frac{{(\Theta_i\xi_j-{\dot{U}}_i)}^2}{2},
\end{equation}
where ${\dot{U}}_i$ is the time derivative of the $i$th state observation, $\Theta_i$ is the $i$th row of the common library $\Theta$ that represents all candidate function values for the ith data. The sparsity constraints mean that the coefficient vector $\xi_j$ is sparse with only a few non-zero entries, each representing a subsistent item in the function library. Subsequently, we exploit the follow-the-regularized-leader (FTRL)-Proximal \cite{34,35,36} style methodology to optimize the outcome by considering solely one instance each time.

Leveraging the fact that each instance is considered individually, we rewrite the loss function and define a single loss term (see Eq. (\ref{eq5})). For example, if the first state measurement $u_1$ is available, the loss is defined as: $L_1\left(\xi_j\right)+\lambda_1\|\xi_j\|_1=0.5(\Theta_1\xi_j-{\dot{U}}_1)^2+\lambda_1\|\xi_j\|_1$. Next, if the second instance of $u_2$ arrives, the loss is defined as: $L_1\left(\xi_j\right)+L_2\left(\xi_j\right)+\lambda_1\|\xi_j\|_1=0.5(\Theta_1\xi_j-{\dot{U}}_1)^2+0.5(\Theta_2\xi_j-{\dot{U}}_2)^2+\lambda_1\|\xi_j\|_1$. In this way, the computer only needs to store information about a single example, thus relieving memory stress. On the $i$th sample, the gradient is calculated as follows:
\begin{equation}
\label{eq6}
{\nabla L}_i\left(\xi_j\right)=(\Theta_i\xi_j-{\dot{U}}_i)\Theta_i.
\end{equation}
Normally, the online gradient descent algorithm can be used to update the coefficient vector $\xi_j^i$ after the arrival of the $i$th data by using:
\begin{equation}
\label{eq7}
{\xi_j^{i+1}=\xi_j^i+C\nabla L}_i\left(\xi_j\right).
\end{equation}

However, this method has been proven to lack general applicability \cite{35}. Correspondingly, the FTRL-proximal style approach is introduced to solve the online problem. We use the following equation to update the coefficient vector $\xi_j^i$:
\begin{eqnarray}
\label{eq8}
\xi_j^{i+1}=&&arg\min_{\xi_j}\sum_{k=1}^{i}{\nabla L}_k\left(\xi_j^k\right)\xi_j+\lambda_1\|\xi_j\|_1\nonumber\\
&&+0.5\lambda_2\|\xi_j\|_2^2+0.5\sum_{k=1}^i\sigma_k\|\xi_j-\xi_j^k\|_2^2,
\end{eqnarray}
where $\lambda_1$ and $\lambda_2$ both denote the regularization coefficient, which is the positive constant, while $\sigma_k$ is on the $i$th learning rate $\eta_i$ aspect, defined by:
\begin{equation}
\label{eq9}
\sum_{k=1}^{i}{\sigma_k=\sqrt t=\eta_i^{-1}}.
\end{equation}

Given the gradient ${\nabla L}_i\left(\xi_j\right)$ a shorthand $g_i$, we set $g_{1:i}=\sum_{k=1}^{i}g_k$. Then the update in Eq. (\ref{eq8}) can be rewritten as follows:
\begin{eqnarray}
\label{eq10}
\xi_j^{i+1}=&&arg\min_{\xi_j}\left(g_{1:i}-\sum_{k=1}^i\sigma_k\xi_j^k\right)\xi_j+\lambda_1\|\xi_j\|_1\\
&&+0.5\left(\lambda_2+\sum_{k=1}^i\sigma_k\right)\|\xi_j\|_2^2+0.5\sum_{k=1}^i\sigma_k\|\xi_j^k\|_2^2,\nonumber
\end{eqnarray}
where the last term $0.5\sum_{k=1}^i\sigma_k\|\xi_j^k\|_2^2$ is a constant with respect to $\xi_j$ and negligibly impacts the update process. Let $Z_i=g_{1:i}-\sum_{k=1}^{i}{\sigma_k\xi_j^k}=Z_{i-1}+g_i-\sigma_i\xi_j^i$, and we ignore the last term and rewrite Eq. (\ref{eq10}) as:
\begin{equation}
\label{eq11}
arg\min_{\xi_j}\left(Z_i\xi_j+\lambda_1\|\xi_j\|_1+0.5\left(\lambda_2+\sum_{k=1}^i\sigma_k\right)\|\xi_j\|_2^2\right).
\end{equation}
Let $\xi_{js}$ and $Z_{i,s} (s=1,2,\ldots,p)$ represent the $s$th element of the vector $\xi_j$ and $Z_i$, respectively. 
Equation (9) is generalized as
\begin{eqnarray}
\label{eq12}
\xi_{js}^{i+1}=&&arg\min_{\xi_{js}}Z_{i,s}\xi_{js}+\lambda_1\left|\xi_{js}\right|\nonumber\\
&&+0.5\left(\lambda_2+\sum_{k=1}^{i}\sigma_k\right)\xi_{js}^2.
\end{eqnarray}

Aiming to simplify the expression, we suppose $w^\ast$ to be the optimal solution of $\xi_{js}^i$ and $\Phi \in \nabla w^\ast$ to be its gradient. Then, Eq. (\ref{eq13}) is satisfied. Equation (\ref{eq14}) shows the subdifferential of $w = sgn(w)$.
\begin{equation}
\label{eq13}
Z_{i,s}+\lambda_1\Phi+(\lambda_2+\sum_{k=1}^i\sigma_k)w^\ast=0.
\end{equation}
\begin{eqnarray}
sgn(w) = 
\left\{
\begin{array}{r}
\Phi \in R|-1 \leq \Phi\leq 1, if\ \ w=0,  \\
1 \ \ \ if \ \ w>0\\
-1,   \ \ \  if \ \ w<0
\end{array}\right.
\label{eq14}
\end{eqnarray}

The solution to $w^\ast$ can be discussed in three cases \cite{39}:
\begin{enumerate}
    \item If $|Z_{i,s}| \leq \lambda_1$,
    \begin{enumerate}
        \item If $w^\ast = 0$, then $sgn(0)\in(-1,1)$ and Eq. (\ref{eq13}) is satisfied.
        \item If $w^\ast > 0$, then $Z_{i,s} + \lambda_1sgn(w^\ast) = Z_{i,s} + \lambda_1 \geq 0$ and $(\lambda_2 + \sum_{k=1}^i \sigma_k)w^\ast > 0$. Equation (\ref{eq13}) does not hold.
        \item If $w^\ast < 0$, then $Z_{i,s} + \lambda_1sgn(w^\ast) = Z_{i,s} -\lambda_1 \leq 0$ and $(\lambda_2 + \sum_{k=1}^i \sigma_k)w^\ast < 0$. Equation (\ref{eq13}) does not hold.
    \end{enumerate}
    
    \item If $|Z_{i,s}| < -\lambda_1$,
    \begin{enumerate}
        \item If $w^\ast = 0$, then $sgn(0) \in (-1,1)$ and Eq. (\ref{eq13}) does not hold.
        \item If $w^\ast > 0$, then $Z_{i,s} + \lambda_1sgn(w^\ast) = Z_{i,s} + \lambda_1 < 0$ and $(\lambda_2 + \sum_{k=1}^i \sigma_k)w^\ast > 0$. Equation (\ref{eq13}) holds and the solution is:
 \begin{equation}
 \label{eq15}
     w^\ast=-\left(\lambda_2+\sum_{k=1}^{i}\sigma_k\right)^{-1}\left(Z_{i,s}+\lambda_1\right).
 \end{equation}
        \item If $w^\ast < 0$, then $Z_{i,s} + \lambda_1sgn(w^\ast) = Z_{i,s} -\lambda_1 < 0$ and $(\lambda_2 + \sum_{k=1}^i \sigma_k)w^\ast < 0$. Equation (\ref{eq13}) does not hold.
    \end{enumerate}

    \item If $|Z_{i, s}| > \lambda_1$,
    \begin{enumerate}
        \item If $w^\ast = 0$, then $sgn(0) \in (-1,1)$ and Eq. (\ref{eq13}) does not hold.
        \item If $w^\ast > 0$, then $Z_{i,s} + \lambda_1sgn(w^\ast) = Z_{i,s} + \lambda_1 > 0$ and $(\lambda_2 + \sum_{k=1}^i \sigma_k)w^\ast > 0$. Equation (\ref{eq13}) does not hold.
        \item If $w^\ast < 0$, then $Z_{i,s} + \lambda_1sgn(w^\ast) = Z_{i,s} -\lambda_1 > 0$ and $(\lambda_2 + \sum_{k=1}^i \sigma_k)w^\ast < 0$. Equation (\ref{eq13}) holds and the solution is:
 \begin{equation}
  \label{eq16}
w^\ast=-\left(\lambda_2+\sum_{k=1}^{i}\sigma_k\right)^{-1}\left(Z_{i,s}-\lambda_1\right).
 \end{equation}
    \end{enumerate}
\end{enumerate}

\begin{algorithm}[t]
\caption{O-SINDy}
\KwIn{$\alpha$, $\beta$, $\lambda_1$, $\lambda_2$: the parameters of O-SINDy\;
    $thresh$: the threshold for truncation\;
    $epo$: the iterations for reusing the finite data\;
    {$\Theta$, $u_t$}: the pre-processed training data\;}
\KwOut{$\xi_j$\;}
   Set $N$ as the length of training data and $p$ as the column number in $\Theta$\;
   \For{each term $s$ \textbf{in} {1: 1: $p$}}{Set $Z_s$=0, $\eta_s$=0, and $\xi_{js}$=0\;}
   \For{each epoch \textbf{in} {1: 1: $epo$}}{
    \For{each instance $i$ \textbf{in} {1: 1: $N$}}{
   Receive instance pair ($\Theta_i, {\dot{U}}_i$) and set $I=\left\{k\ \left|\Theta_{ik}\neq0\right.\right\}$\;
   \For{each element $k$ \textbf{in} $I$}{
   Compute $\xi_{jk}^i$ using  Eq. (\ref{eq20})\;
    \For{each element $k$ \textbf{in} $I$}{
   $g=(\Theta_i\xi_j^i-{\dot{U}}_i)\Theta_{ik}$\;
   $\sigma=(\sqrt{\eta_k+g^2}-\sqrt{\eta_k})/\alpha$\;
   $Z_k=Z_k+g-\sigma\xi_{jk}^i$\;
   $\eta_k=\eta_k+g^2$\;
   }}}}
   \For{each term $s$ \textbf{in} {1: 1: $p$}}{
   Set $\xi_{js}=0$ if $\left|\xi_{js}\right|<thresh$\;
   }
\end{algorithm}

The above discussion in different estimation scenarios is summarized in Eq. (\ref{eq17}), where $w^\ast=\xi_{js}^i$.
\begin{eqnarray}
 \label{eq17}
w^\ast = 
\left\{
\begin{array}{r}
0, \ \ |Z_{i,s}|\leq \lambda_1  \\
-\frac{\left(Z_{i,s}-\lambda_1sgn\left(Z_{i,s}\right)\right)}{\left(\lambda_2+\sum_{k=1}^{i}\sigma_k\right)}, \ \ otherwise
\end{array}\right.
\end{eqnarray}

Moreover, we set a remarkable learning rate
 \begin{equation}
 \label{eq18}
\eta_{is}=\frac{\alpha}{(\beta+\sqrt{\sum_{k=1}^{i}{(g_{k,s})}^2})},
 \end{equation}
where $\alpha$ and $\beta$ are both positive constants, $g_{k,s}$ denotes the $s$th element of the gradient $g_k$ and $\eta_{is}$ is the learning rate of $g_{k,s}$. Thus, Eq. (\ref{eq9}) can be rewritten as follows:
 \begin{equation}
 \label{eq19}
\sum_{k=1}^{i}\sigma_k=\frac{1}{\eta_{is}}=\frac{(\beta+\sqrt{\sum_{k=1}^{i}{(g_{k,s})}^2})}{\alpha}.
 \end{equation}

The final form of the closed solution is obtained and expressed as 
\begin{eqnarray}
\xi_{js}^i = 
\left\{
\begin{array}{r}
0, \ \ |Z_{i,s}|\leq \lambda_1  \\
\frac{(\lambda_1sgn\left(Z_{i,s}\right)-Z_{i,s})}{(\lambda_2+\frac{(\beta+\sqrt{\sum_{k=1}^{i}{(g_{k,s})}^2})}{\alpha})}, \ \ otherwise,
\end{array}\right.
\label{eq20}
\end{eqnarray}
where $\alpha$ and $\beta$ are both positive constants, $g_{k,s}$ denotes the $s$th element of the gradient $g_k$.

Notably, $\xi_{js}$ is always updated according to the current input instance, thereby guiding the detection of change points if there exists any variation in the evolving system. Additionally, the mechanism frees the computer from storing the whole large-scale data. It should be highlighted that the method suggested above is entirely different from the batch learning methods, which update gradients depending on all available measurements. O-SINDy \footnote{Available code: https://github.com/xiaoyuans/O-SINDy.} takes advantage of a single available instance and updates gradients based on the current loss $L_i\left(\xi_j\right)(i=1,2,\ldots,mn)$ in each iteration, thereby being easily extended to handle evolving systems. Note that the optimum selection of hyper-parameters in each case is determined by grid search ($10^i, i =-8, -7, …, 1$).

\begin{table*}[htbp]
\caption{\label{tab:table2}%
Summary of online identification results for a wide range of canonical models. O-SINDy is applied to reconstruct the correct model structure in each example. Settings for the spatial and temporal sampling of the numerical simulation data is given, along with the relative $L_2$ error in recovering the parameters of these dynamical models.}
\begin{ruledtabular}
\renewcommand\arraystretch{1.15}
\begin{tabular}{p{3cm}<{\centering}p{7cm}p{1.5cm}<{\centering}<{\centering}p{5cm}<{\centering}}
Dynamic System & Identified Governing Equation&	Error &	Setting\\
\colrule
Harmonic oscillator1 & $\dot{x}=-0.1000x+2.0000y$ & 1.62e-05 &	$\left[x_0,y_0\right]=\left[2,0\right]$,\\
& $\dot{y}=-2.0000x-0.1000y$& &$m=2500$\\
\colrule
Harmonic oscillator2 & $\dot{x}=-0.0998x^3+2.0014y^3$ & 6.28e-04 &	$\left[x_0,y_0\right]=\left[2,0\right]$,\\
& $\dot{y}=-2.0011x^3-0.1001y^3$ & &$m=2500$\\
\colrule
Lorenz system 1	 & $\dot{x}=-9.9990x-10.8988y$
&	2.80e-04	&$\left[x_0,y_0,z_0\right]=\left[-8,7,27\right],
$\\
& $\dot{y}=-0.8999y+9.6000xz$&&$m=10000$\\
&$\dot{z}=-7.1000-2.6998z-3.1000xy$&&\\
\colrule
Lorenz system 2 & $\dot{x}=5.7004z-3.5007xy+2.1001yz$
&	2.70e-04	&$\left[x_0,y_0,z_0\right]=\left[-8,7,27\right],
$\\
& $\dot{y}=-10.3001-2.6998y+2.5999x^2-z^2$&&$m=10000$\\
&$\dot{z}=-10.8828z-5.5957xy+3.3992yz$&&\\
\colrule
Hopf normal form & $\dot{x}=0.2514x-0.9995y-{1.0060x}^3-1.0009xy^2$ & 1.85e-02	& $\left[x_0,y_0\right]=\left[2,0\right]$,\\
& $\dot{y}=1.0006x+0.2606y-{1.0422x}^2y-1.0392y^3$& &$m=3000, \mu=0.25, \omega=1, A=1$\\
\colrule
Diffusion from random walk	& $u_t=0.5000u_{xx}$ & 1.69e-05 &	$t\in\left[0,0.02\right]$, $m=3, n=8000$\\
\colrule
Burgers’ &	$u_t=-0.9993uu_x+0.1002u_{xx}$ & 6.97e-04	& $t\in\left[0,10\right], m=101$
$x\in\left[-8,8\right], n=256$\\
\colrule
Korteweg-de Vries (KdV)	& $u_t=-6.1317uu_x-1.0029u_{xxx}$& 2.17e-02 &	$t\in\left[0.05,0.175\right],m=6$, 
$x\in\left[-10,12\right], n=256$\\
\colrule
Kuramoto-Sivashinsky & $u_t=-0.9667uu_x-0.9585u_{xx}-0.9646u_{xxxx}$&  2.55e-02	& $t\in\left[0,100\right], m=251$,
$x\in\left[0,100\right], n=1024$\\
\colrule
Reaction-diffusion	& $u_t=0.1000u_{xx}+0.1001u_{yy}-1.0001uv^2-1.0001u^3+0.9996v^3+0.9996u^2v+1.0001u$& 2.65e-04	& $t\in\left[0,10\right],m=201$, $subsample\ 0.285$,\\
& $v_t=0.1001v_{xx}+0.1001v_{yy}+1.0001v-0.9996uv^2-{0.9996u}^3-1.0001v^3-1.0001u^2v$ & & $x,y\in\left[-10,10\right],n=512$\\
\end{tabular}
\end{ruledtabular}
\end{table*}

\section{\label{sec:level3}Results}
\subsection{\label{sec:level31}Discovering Single System from Streaming Data}

\subsubsection{\label{sec:level311}Example – Discovering Chaotic Lorenz System}

The online algorithmic procedure for identifying correct governing differential equations of the chaotic Lorenz system from time series is demonstrated in Fig.~\ref{fig1}. Our proposal combines data collection, a library of potential candidate functions, and the online reconstruction method. The Lorenz system is highly nonlinear described by the following equations, according to which we construct a canonical dynamic instance for model discovery.
\begin{eqnarray}
\dot{x}=\sigma\left(y-x\right),
\\
\dot{y}=x\left(\rho-z\right)-y,
\\
\dot{z}=xy-\beta z.
\label{eq21}
\end{eqnarray}

To indicate the nonlinear combination of Lorenz variables, common parameters are $\sigma=10$, $\beta$=8/3, and $\rho$=28, with the initial condition of $(x_0, y_0, z_0)^T = (-8, 7, 27)^T$. Ref. \cite{3} has employed the SINDy autoencoder to discover a parsimonious model with only seven active terms that seems to mismatch the original Lorenz system. Interestingly, the dynamics of the resulting model exhibit an attractor with a 2-lobe structure, which is qualitatively consistent with the true Lorentz attractor. Then the sparsity pattern can be rewritten in the standard form by choosing an appropriate variable transformation. Considering the chaotic nature, O-SINDy show the ability to capture the concise model on two Lorenz systems. Results in Fig.~\ref{fig1}(d) and Table~\ref{tab:table2} imply that O-SINDy can accurately reproduce the attractor dynamics from chaotic trajectory measurements. In both cases, we conducted 10000 timesteps with time intervals $dt$=0.01.

Given time intervals and initial conditions, we simulate the Lorenz system within a certain time horizon. Specifically, state measurements are collected under the constraints of the transformed expressions, including historical data of the state $U$ and its time derivative $U_t$. The combination of spatial and temporal modes results in the trajectory of dynamics shown in Fig.~\ref{fig1}(a). A library of potential candidate functions, $\Theta\left(U\right)$, is constructed to find the least terms required to satisfy $U_t=\Theta\left(U\right)\Xi$. Candidate functions can be polynomial functions, trigonometric functions, exponential functions, logarithmic functions, partial derivatives, constant items, and any additional terms about $U$. The proposed online technique can identify the coefficient $\Xi=[\xi_1;\xi_2;\ldots;\xi_d]$ to determine active terms of the dynamics. To elude the obstacle of large-scale data storage and batch gradient update, we design a cumulative loss function to reflect the mode of streaming data by solely analyzing the information of one sample at a time. The pre-established sparse scenario guarantees that the resulting solution obtained by the online method can effectively balance model complexity with description ability to avoid overfitting, thereby promoting its interpretability and extensibility. The prompt update of components and coefficients in the governing equation is one implementation of data arrival based synchronization. Consequently, we can subtly detect moderate or drastic variations in the system through changes in the resulting model.

\begin{figure*}[htbp]
\includegraphics[width=0.8\linewidth]{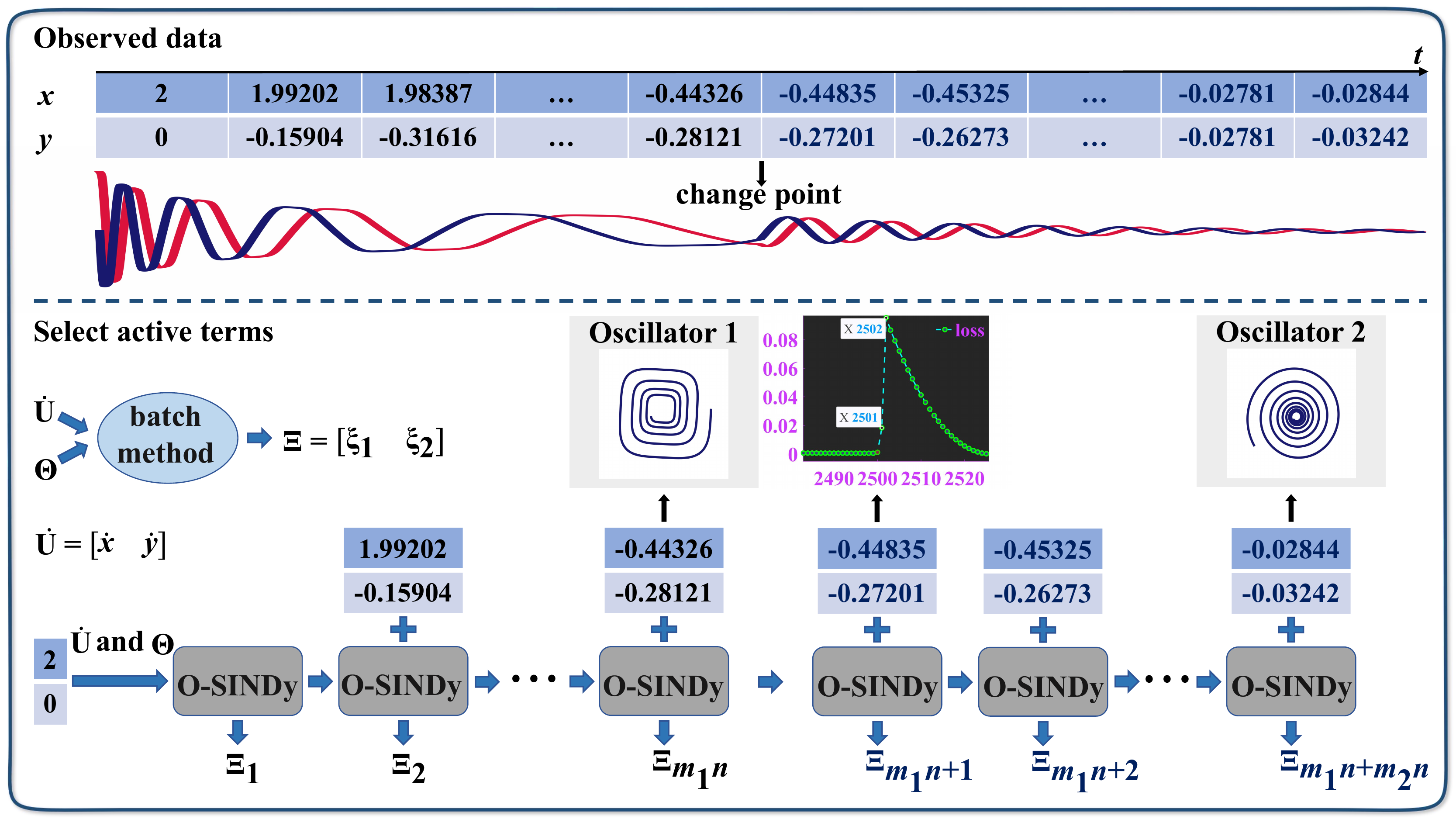}
\caption{Difference between batch learning methods and O-SINDy in reconstructing governing equations from evolving systems. As instances successively arrive, the change point arises when the system is disturbed, and the original distribution of measurements varies. The batch method can only strike a compromise solution utilizing all data generated by the two completely different systems. By means of O-SINDy, the loss value gradually decreases and stabilizes unless the change point appears. The former governing equation is reconstructed right before the change point, and the latter is identified after a few iterations.}
\label{fig3}
\end{figure*}

\subsubsection{\label{sec:level34}Performance on Discovery for Canonical Models}

We perform O-SINDy on ten canonical models in mathematical physics and engineering sciences, and the results are shown in Table ~\ref{tab:table2}. These physical systems contain dynamics with periodic to chaotic behavior, ranging from linear to strongly nonlinear systems. The settings for spatial and temporal sampling and the coefficient errors are detailed in Table~\ref{tab:table2}. Encouragingly, O-SINDy can recover every physical system, even those with significant spatial subsampling. The notable results highlight the broad applicability of the method and the success of this online technique in discovering governing differential equations. Remarkably, the capacity for capturing nontrivial active terms, in particular, has essential explanatory implications for model discovery.

\begin{table*}[htbp]
\caption{\label{tab:table3}%
Summary of O-SINDy and standard machine learning methods for identifying the evolving systems. O-SINDy is successfully applied to reconstruct the correct governing PDEs before and after the system changes, while the standard machine learning methods strike a compromise solution utilizing all samples.}
\begin{ruledtabular}
\centering
\renewcommand\arraystretch{1.15}
\begin{tabular}{cp{6.5cm}p{6.5cm}}
Methodology &	Cubic\_Linear & KdV\_KS\\
\colrule
O-SINDy	& \textbf{Phase 1}	& \textbf{Phase 1}\\
& $\dot{x}=-0.1257x^3+2.0006y^3$ & $u_t=-6.1036uu_x-0.9460u_{xxx} $\\
& $\dot{y}=-2.0193x^3-0.1179y^3$ & \\
& \textbf{Phase 2}	& \textbf{Phase 2}\\
 &	$\dot{x}=-0.1006x+1.9958y$ & $u_t=-0.9894uu_x-0.9713u_{xx}-0.9158u_{xxxx}$\\
& $\dot{y}=-1.9999x-0.1000y$ & \\
\colrule
STRidge \cite{4} &	$\dot{x}=0.5366y+1.3167y^3-0.4861x^2y+0.1735y^5+0.1678x^2y^3+0.1170x^4y$ & $u_t=-1.6007u_x+1.4328uu_{xx}-0.5409u^2u_{xx}-0.5814u_{xxx}+0.8952uu_{xxx}+0.5741uu_{xxxx}$\\
& $\dot{y}=-0.5751x+0.5912xy^2+0.1335x^2y-1.3422x^3-0.1340xy^4-0.2122x^3y^2-0.1503x^5 $ & \\
\colrule
TrainSTRidge \cite{4}	& $\dot{x}=0.5340y+1.3186y^3-0.4700x^2y+0.1734y^5+0.1610x^2y^3+0.1114x^4y$	& $u_t=-1.6007u_x+1.4328uu_{xx}-0.5409u^2u_{xx}-0.5814u_{xxx}+0.8952uu_{xxx}+0.5741uu_{xxxx} $\\
&$\dot{y}=-0.5751x+0.5912xy^2+0.1335x^2y-1.3422x^3-0.1340xy^4-0.2122x^3y^2-0.1503x^5 $&\\
\colrule
STLS \cite{42} &	$\dot{x}=0.4920y+1.4390y^3-0.4630x^2y+0.1310y^5+0.1120x^2y^3+0.1320x^4y$ 	& $u_t=-1.6007u_x+1.4328uu_{xx}-0.5409u^2u_{xx}-0.5814u_{xxx}+0.8952uu_{xxx}+0.5741uu_{xxxx} $\\
& $\dot{y}=-0.5260x+0.4480xy^2-0.1320x^2y-1.4280x^3-0.0830xy^4-0.1510x^3y^2-0.1240x^5$ &\\
\colrule
Lasso \cite{14}	& $\dot{x}=0.5367y+1.3166y^3-0.4872x^2y+0.1735y^5+0.1681x^2y^3+0.1179x^4y$ & $u_t=-1.6007u_x+1.4328uu_{xx}-0.5409u^2u_{xx}-0.5814u_{xxx}+0.8952uu_{xxx}+0.5741uu_{xxxx} $\\
&$\dot{y}=-0.5748x+0.5925xy^2+0.1340x^2y-1.3340x^3-0.1339xy^4-0.2136x^3y^2-0.1495x^5$&\\
\colrule
ElasticNet \cite{43} & $\dot{x}=0.4364y+1.3402y^3+0.1751y^5$ & $u_t=-1.6007u_x+1.4328uu_{xx}-0.5409u^2u_{xx}-$\\
&$\dot{y}=-0.4410x-0.0945y^3-1.4026x^3-0.1417x^5$&$0.5814u_{xxx}+0.8952uu_{xxx}+0.5741uu_{xxxx} $\\
\colrule
FoBaGreedy \cite{44} & $\dot{x}=0.2302y+1.8805y^3-0.1003x^3$ & $u_t=-1.5771u_x+1.5007uu_{xx}-0.5332u^2u_{xx}-$\\
&$\dot{y}=-0.2487x-0.0988y^3-1.8757x^3$&$0.6168u_{xxx}+0.9653uu_{xxx}$\\
\end{tabular}
\end{ruledtabular}
\end{table*}

\subsection{\label{sec:level35}Discovering Hybrid Systems}

We first consider an evolving two-dimensional damped harmonic oscillator that changes from cubic to linear.
\begin{equation}
\frac{d\left[\begin{matrix}x\\y\\\end{matrix}\right]}{dt}=\left[\begin{matrix}-0.1&2\\-2&-0.1\\\end{matrix}\right]\left[\begin{matrix}x\\y\\\end{matrix}\right],
\label{eq26}
\end{equation}
\begin{equation}
\frac{d\left[\begin{matrix}x\\y\\\end{matrix}\right]}{dt}=\left[\begin{matrix}-0.1&2\\-2&-0.1\\\end{matrix}\right]\left[\begin{matrix}x^3\\y^3\\\end{matrix}\right].
\label{eq27}
\end{equation}
A particle in simple harmonic vibration is known as a harmonic oscillator, whose motion is the simplest ideal vibration model. Fig.~\ref{fig3} illustrates the dynamic data and the detection of change points.
Firstly, we create the solution to Eq.~(\ref{eq27}), $U_1$, with 2,500 timesteps and the initial data $[2,0]$. Then, we utilize the last instance in $U_1$ as the initial data and create the solution to Eq.~(\ref{eq26}) with 2,500 timesteps as well, recorded as $U_2$. Specifically, the change point arises when the system changes, and then the original distribution of generated data varies. Therefore, the corresponding loss value gradually decreases and stabilizes unless the change point appears. In this context, the coefficients in the former governing equation and the change location are simultaneously identified according to the sudden increase of loss value.

With an augmented nonlinear library including polynomials up to the fifth order, the change point is spotted in evolving circumstances. The loss curve tends to stabilize faster after resetting the training coefficient, thereby accelerating the convergence of O-SINDy because the former result may mislead the training process.

\begin{figure}[htbp]
\includegraphics[width=\linewidth]{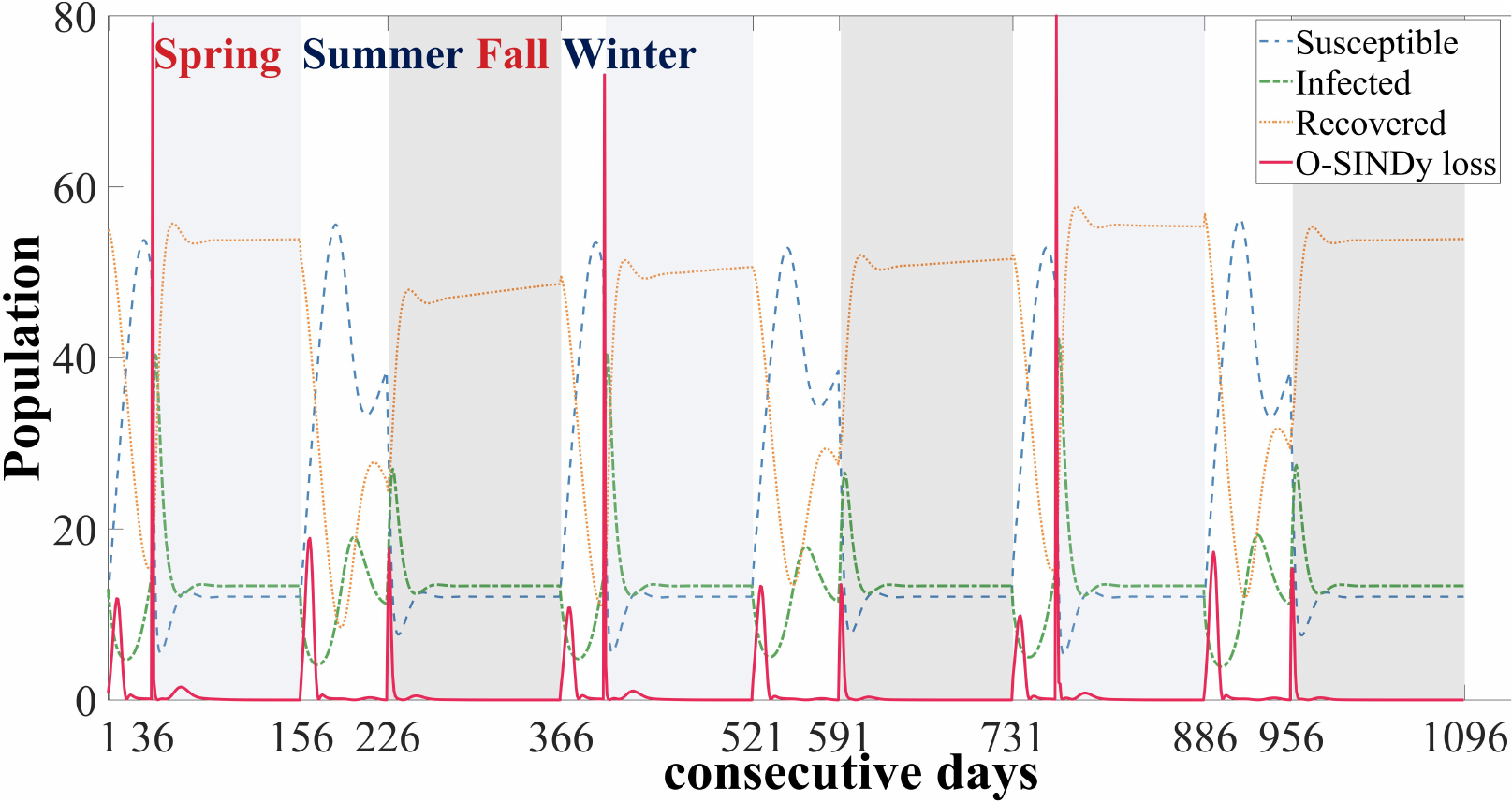}
\caption{The results of O-SINDy in the SIR disease model with varying transmission rates. The infected population dynamics over 3 years show declines in summer and winter, followed by outbreaks in spring and fall. The loss peak in O-SINDy appears with the varying $\beta\left(t\right)$.}
\label{fig5}
\end{figure}

The transition from the KdV to the Kuramoto-Sivashinsky equation is simulated as an illustrative example that exhibits qualitatively different dynamics as the system evolves. Summary of O-SINDy and the state-of-the-art methods in SINDy for this identification in Table~\ref{tab:table3} demonstrates that our approach works well. Most data-driven methods regard batch measurements as a whole and update gradients integrally to obtain an ephemeral fitting. Owing to the ignorance of variations, existing methodologies are biased to strike a compromise solution, so that strenuous to identify the changes from consecutively generated data. Quite the contrary, O-SINDy succeeds in simultaneously estimating forms and identifying changes of the governing equations.

Moreover, we investigate the Susceptible-Infected-Recovered (SIR) disease model with time-varying transmission rates, widely studied in epidemiology \cite{grenfell1994measles,mangan2019model}. The following equation is a mathematical description of the SIR model.
\begin{eqnarray}
& \dot{S}=vN-\frac{\beta\left(t\right)}NIS-dS,
\\
&\dot{I}=\frac{\beta\left(t\right)}NIS-\left(\gamma+d\right)I,
\\
&\dot{R}=\gamma I-dR,
\\
&\beta\left(t\right) = 
\left\{
\begin{array}{r}
\hat{\beta}\left(1+b\right),\ \ t\in Spring / Fall,\\
\hat{\beta}\frac{1}{1+b},\ \ t\in Summer / Winter.
\end{array}\right.
\label{eq28}
\end{eqnarray}
It is a time-dependent hybrid dynamical system, where $v=d=0.0027$, $N=1000$ is the total population, $\gamma=0.2$, $b=0.8$, and $\hat{\beta}=9.336$ is a base transmission rate. We simulate the model for three years within the initial condition at $[S_0, I_0, R_0]=[12, 13, 975]$, recording at a daily interval. A random perturbation is added to the start of each season with the same probability.

To illustrate the dynamics of the SIR model, we display the recovered population by subtracting 920 and scaling the loss in O-SINDy by the maximum value. As shown in Fig.~\ref{fig5}, O-SINDy captures the transition between states by the dramatically increasing loss. In this example, the perturbation is able to be identified utilizing solely the information of the recovered population, which is ignored in Hybrid-Sparse Identification of Nonlinear Dynamics (Hybrid-SINDy)\cite{mangan2019model}. It is an offline batch learning method and outputs the governing equation of a single day by repeatedly clustering the whole time-series data during the model selection step. The identified governing equations of four random days in the four seasons are summarized in Table~\ref{tab:table4}. The results show that O-SINDy provides more precise reconstruction accuracy for a single day. Specifically, instances of the same coefficients share the hyper-parameters in O-SINDy. At the same time, Hybrid-SINDy selects a model of the highest frequency from results recovered by SINDy with different hyper-parameters for each certain instance. Moreover, the effect of cluster size is particularly vital to solutions in Hybrid-SINDy.
\begin{table*}[htbp]
\caption{
The results of O-SINDy and Hybrid-SINDy in identifying the SIR model of four random days. O-SINDy provides better reconstructions for a single day, while Hybrid-SINDy misses some significant terms.}
\begin{ruledtabular}
\centering
\renewcommand\arraystretch{1.0}
\begin{tabular}{cp{5cm}p{5cm}p{5cm}}
  Datetime & True models & O-SINDy & Hybrid-SINDy \\
\colrule
30 & $\dot{S}=2.7397-0.0052IS-0.0027S$ &	$\dot{S}=2.7295-0.0052IS-0.0024S$ &  $\dot{S}=2.7397-0.0052IS-0.0027S$\\
$\left(Spring\right)$ & $\dot{I}=0.0052IS-0.2027I$ & $\dot{I}=0.0052IS-0.2028I$ & $\dot{I}=0.0730I$\\
\colrule
80 & $\dot{S}=2.7397-0.0168IS-0.0027S$ &	$\dot{S}=2.7397-0.0168IS-0.0027S$ & $\dot{S}=2.7243-0.0169IS$\\
$\left(Summer\right)$ & $\dot{I}=0.0168IS-0.2027I$ & $\dot{I}=0.0168IS-0.2027I$ & $\dot{I}=0.0168IS-0.2027I$\\
\colrule
184 & $\dot{S}=2.7397-0.0052IS-0.0027S$ &	$\dot{S}=2.7064-0.0052IS+0.0040I$ & $\dot{S}=2.7397-0.0052IS -0.0027S$\\
$\left(Fall\right)$ & $\dot{I}=0.0052IS-0.2027I$ & $\dot{I}=-0.0038+0.0052IS-0.2023I $ & $\dot{I}=0.0777I$\\
\colrule
350 & $\dot{S}=2.7397-0.0168IS-0.0027S$ &	$\dot{S}=2.7397-0.0168IS-0.0027S$ & $\dot{S}=0.2249S+0.2054I-0.0339IS$\\
$\left(Winter\right)$ & $\dot{I}=0.0168IS-0.2027I$ & $\dot{I}=0.0168IS-0.2027I $ & $\dot{I}=0.0168IS-0.2027I$\\
\end{tabular}
\end{ruledtabular}
\label{tab:table4}
\end{table*}

\subsection{\label{sec:level36}Discovering Switching Linear Systems}
We also test O-SINDy on two switching linear systems, including synthetic and real-world datasets. Define $x_t \in R^N$ as the sampled trajectory of a dynamical system, $N$ is the feature dimension. Time-varying autoregressive model with low-rank tensors (TVART) \cite{harris2021time} holds the assumption, $x_{t+1}=A_tx_t$, where $A_t$ is constant within a time window but varies in the switching system. On the contrary, O-SINDy discards the time window and admits that $A_t$ may change at every time point.

\begin{figure*}[htbp] 
\includegraphics[width=0.8\linewidth]{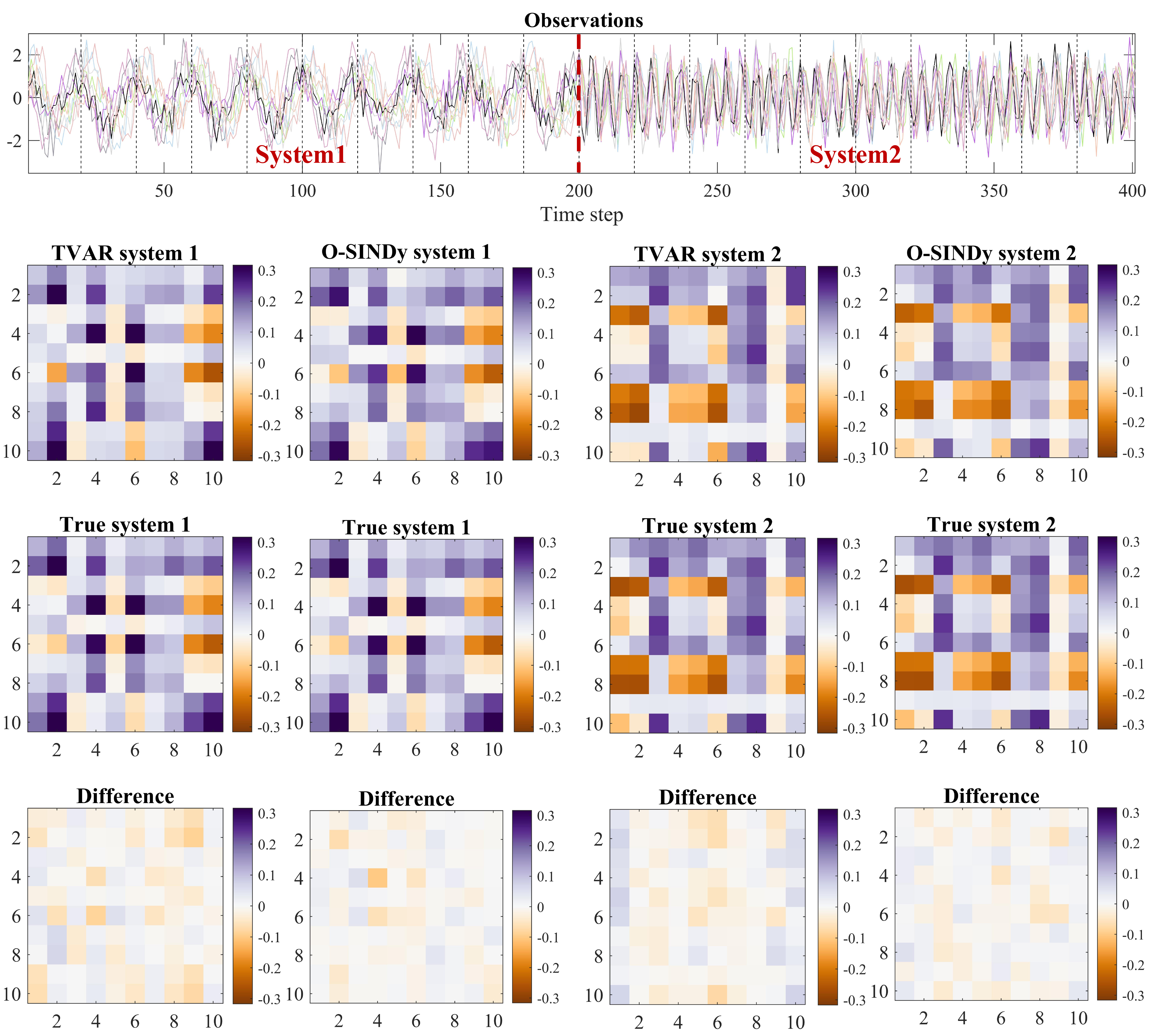}
\caption{The results of O-SINDy and TVART in synthetic switching linear systems. The difference between identified and true system matrices is slight in O-SINDy.}
\label{fig6}
\end{figure*}

Regarding the synthetic example, we randomly generate two system matrices $A_{N\times N}^1$ and $A_{N\times N}^2$ to encourage the switching system. The dynamics follow $A_{N\times N}^1$ and switch to $A_{N\times N}^2$ after half of the time series. Results illustrated in Fig.~\ref{fig6} show that O-SINDy is competitive with the state-of-the-art technique TVART.

\begin{figure}[htbp]
\includegraphics[width=\linewidth]{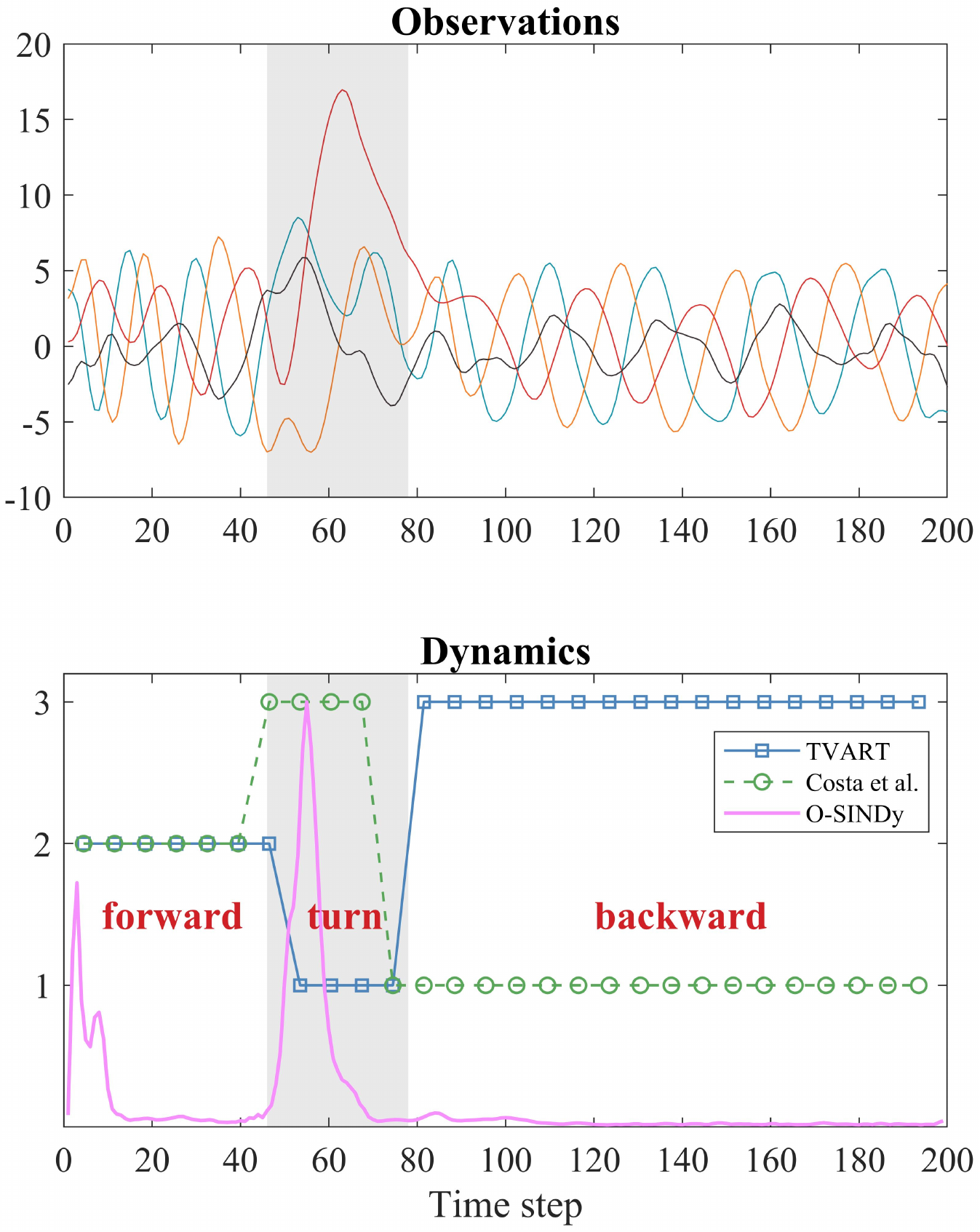}
\caption{The worm's posture dynamics of O-SINDy and compared methods.}
\label{fig7}
\end{figure}

The posture dynamics of the worm Caenorhabditis elegans \cite{stephens2008dimensionality,broekmans2016resolving} is studied as the real-world example for clustering via adaptive, locally linear models \cite{costa2019adaptive}. This time-series data details the escape behavior of the worm in response to a heat stimulus, e.g., the transition from forward, turn, to backward crawling. Fig.~\ref{fig7} characterizes the true dynamics and the discovered modes corresponding to its three typical dynamical regimes. It should be highlighted that the bifurcation of the worm dynamics is interpretable on a smaller time scale with the application of O-SINDy.

\section{\label{sec:level4}Conclusion and Discussion}

We have proposed an online modeling method, O-SINDy, capable of finding governing differential equations of evolving systems from streaming data.
We demonstrate that O-SINDy works on various canonical instances. Results in Table~\ref{tab:table2} indicate that the inference of governing equation is accurate when utilizing O-SINDy on time series from numerical simulations. Additionally, we show the excellent performance of simultaneously estimating forms and identifying changes in time-varying dynamic systems. 

More sampling points or longer time series correspond to the preferable identification of the internal control structure, while the KdV equation is an exception (see Table~\ref{tab:tables1}). One potential viewpoint is that approximating the soliton solution introduces great uncertainty. Nevertheless, it remains an open question to estimate the required time-series length for distilling the accurate underlying governing differential equations. The implementation of existing methodologies fundamentally depends on sufficiently large datasets, even though the dynamics are only a parsimonious representation. O-SINDy, on the contrary, is demonstrated to recover the true governing equations by iteratively reusing the finite available time series. Note that the computation of time derivatives results in the main error, which is magnified by the numerical roundoff. Thus, correctly estimating numerical derivatives is the most critical and challenging task for O-SINDy, especially in a noisy context.

O-SINDy is a viable tool capable of tackling streaming data from evolving systems for accomplishing the assignment of model discovery. The integration opens up a novel, interesting research insight for real-time modeling, online analysis, and control techniques of complex dynamic systems.
\begin{table}[htbp]
\renewcommand\arraystretch{1.25}
\caption{\label{tab:tables1}%
Summary of O-SINDy for identifying the KdV equations of different spatial and temporal sampling of the numerical simulation data.}
\begin{ruledtabular}
\begin{tabular}{cccccc}
Discretization & $m$=3 & $m$=6	&$m$=10	&$m$=10	&$m$=6\\
	& $n$=256	& $n$=256 &	$n$=256 &	$n$=128 &	$n$=128\\
 \colrule
Error& 	0.2807	& 0.0217 &	0.0090 &	0.0972 &	0.0068\\
\end{tabular}
\end{ruledtabular}
\end{table}

\begin{acknowledgments}
This work was supported in part by the National Natural Science Foundation of China under Grant 62206205, in part by the Fundamental Research Funds for the Central Universities under Grant XJS211905, in part by the Guangdong High-level Innovation Research Institution Project under Grant 2021B0909050008, and in part by the Guangzhou Key Research and Development Program under Grant 202206030003.
\end{acknowledgments}

\appendix
\section{Data Preprocessing}
For massive-scale datasets, sparse sampling can be used to reduce the data size. For identification, it should be noted that we only require a small number of spatial points and their neighbors, whose responsibility is estimating the partial derivative terms in the candidate library. That is, local information around each measurement is necessarily wanted. Distinction allowing for application to subsampled spatiotemporal sequences, to an extent, is critically important due to experimentally and computationally prohibitive implementation of collecting full-state measurements \cite{4}. 
In terms of derivation, second-order finite differences \cite{37} are devoted to the clean data from numerical simulations, while the easiest to implement and most reliable method for the noisy data is a polynomial interpolation \cite{38}. In principle, we abandon those points close to the boundaries due to the absence of numerical derivatives. If we know any prior knowledge about the governing equation, for instance, if one of the potential terms is determined to be nonzero in advance, we can apply the additional information to the initialization phase. Moreover, truncation of the solution is a tool to maintain its sparsity, and different threshold values may provide distinguishing sparsity levels of the final output. As for the artificial noise added to the state measurements of governing equations, we use white noise. The noisy data is directly trained in experiments. In terms of the reaction-diffusion equation, exceptionally, we use the singular value decomposition (SVD) \cite{40} to denoise some noisy spatiotemporal series, and the result is a low-dimensional approximation of datasets.


\section{Canonical Models}
\subsection{Reaction Diffusion}

The dynamic of reaction-diffusion systems is defined by the following equations, which provide great expressiveness and freedom although they are deceptively simple.
\begin{eqnarray}
u_t=0.1\nabla^2u+\lambda\left(A\right)u-w\left(A\right)v,
\\
v_t=0.1\nabla^2v+w\left(A\right)u+\lambda\left(A\right)v.
\label{eq24}
\end{eqnarray}
where $A=u^2+v^2$, $w\left(A\right)=-\beta A^2$, and $\lambda\left(A\right)=1-A^2$. Given the vast number of data points, we subsample 5000 discretized spatial points with 30 time points each.

\subsection{Hopf Norm Form}
In this example, the application of O-SINDy is extended to a parameterized normal form of Poincaré-Andronov-Hopf bifurcation.
\begin{eqnarray}
\dot{x}=\mu x-wy-Ax\left(x^2+y^2\right),
\\
\dot{y}=wx+\mu y-Ay\left(x^2+y^2\right).
\label{eqhnf}
\end{eqnarray}
We explored a normal form of Hopf bifurcation on two-dimensional ordinary differential equations systems. Given $w=1$ and $A=1$, we collected data from the noise-free system for eight various values of the parameter $\mu$.

\subsection{Fokker-Planck Equation}
Fokker-Planck equation that reflects the connection between the diffusion equation and Brownian motion has been taken into account. Considering the simplest form of the diffusion equation where the diffusion coefficient is 0.5 and the drift term is zero, the corresponding Fokker-Planck equation is $u_t=0.5u_{xx}$. The Brownian motion model is usually simplified as a random walk in physics, such as the random movement of molecules in liquids and gases. To simulate the process, we add a distributed random variable with variance $dt=0.01$ to the time series and sample the movement of the random walker.

\subsection{Burgers’ Equation}

Here, we consider a fundamental partial differential equation in the fields of fluid mechanics, nonlinear acoustics, and aerodynamics. As a dissipative system in one-dimensional space, the general form of Burgers’ equation is described by a diffusion coefficient $c$ (also known as kinematic viscosity). The speed of the fluid at the indicated spatial coordinate $x$ and temporal coordinate $t$ in a thin ideal pipe can be expressed as the following equation:
\begin{equation}
    u_t={-uu_x+cu}_{xx},
\end{equation}
where the diffusion term $u_{xx}$ contributes the advective form of the Burgers’ equation.
\subsection{Korteweg-de Vries Equation}
The Korteweg-de Vries (KdV) equation is a nonlinear, partial differential equation for a function $u$ of two real variables, $x$ and $t$, which refer to space and time, respectively. Numerically, its solutions seem to be decomposed into a collection of well-separated solitary waves that are almost unaffected in shape by passing through each other. The soliton solution is given by
\begin{equation}
    u\left(x,t\right)=0.5c\times{sech}^2[0.5\sqrt c(x-ct-a)],
\end{equation}
where $c$ stands for the phase speed, $a$ is an arbitrary constant and $sech\left(x\right)=1/cosh\left(x\right)$ is the hyperbolic secant function. This equation describes a right-moving soliton that propagates with a speed proportional to the amplitude.

The correct equation cannot be distinguished from a single propagating soliton solution, due to the fact that some studied expressions may be solutions to more than one PDE \cite{4}. For example, both the one-way wave equation $u_t + cu_x = 0$ and the KdV equation admit the same traveling wave solution of the form $u = f(x - ct)$ if the initial data was a hyperbolic secant squared. Hence, we constructed time-series data for more than a single initial amplitude to rectify the ambiguity in selecting the governing PDE, thereby enabling the unique determination. 

As the circumstance under a single propagating soliton solution, we respectively constructed two solutions without noise having the traveling speed $c$ equal to 5 or 1 on grids with 6 timesteps and 256 spatial points. Ultimately, corresponding advection equations with different $c$ were identified using a single traveling wave. We believe that two waves with different amplitudes and speeds may solve the KdV equation.

\subsection{Kuramoto-Sivashinsky Equation}

The Kuramoto-Sivashinsky (KS) equation is a fourth-order nonlinear PDE derived by Yoshiki Kuramoto and Gregory Sivashinsky in the late 1970s. Specifically, it provides two dissipative terms $u_{xxx}$ and $u_{xxxx}$ based on Burgers’ equation, where the fourth-order diffusion term accomplishes the stabilizing regularization rather than the second-order diffusion term $u_{xx}$, which leads to long-wavelength instabilities. By leveraging a spectral method, the numerical solution to the KS equation was created with 101 timesteps and 1024 spatial points.

\nocite{*}


\begin{thebibliography}{63}%
\makeatletter
\providecommand \@ifxundefined [1]{%
 \@ifx{#1\undefined}
}%
\providecommand \@ifnum [1]{%
 \ifnum #1\expandafter \@firstoftwo
 \else \expandafter \@secondoftwo
 \fi
}%
\providecommand \@ifx [1]{%
 \ifx #1\expandafter \@firstoftwo
 \else \expandafter \@secondoftwo
 \fi
}%
\providecommand \natexlab [1]{#1}%
\providecommand \enquote  [1]{``#1''}%
\providecommand \bibnamefont  [1]{#1}%
\providecommand \bibfnamefont [1]{#1}%
\providecommand \citenamefont [1]{#1}%
\providecommand \href@noop [0]{\@secondoftwo}%
\providecommand \href [0]{\begingroup \@sanitize@url \@href}%
\providecommand \@href[1]{\@@startlink{#1}\@@href}%
\providecommand \@@href[1]{\endgroup#1\@@endlink}%
\providecommand \@sanitize@url [0]{\catcode `\\12\catcode `\$12\catcode
  `\&12\catcode `\#12\catcode `\^12\catcode `\_12\catcode `\%12\relax}%
\providecommand \@@startlink[1]{}%
\providecommand \@@endlink[0]{}%
\providecommand \url  [0]{\begingroup\@sanitize@url \@url }%
\providecommand \@url [1]{\endgroup\@href {#1}{\urlprefix }}%
\providecommand \urlprefix  [0]{URL }%
\providecommand \Eprint [0]{\href }%
\providecommand \doibase [0]{https://doi.org/}%
\providecommand \selectlanguage [0]{\@gobble}%
\providecommand \bibinfo  [0]{\@secondoftwo}%
\providecommand \bibfield  [0]{\@secondoftwo}%
\providecommand \translation [1]{[#1]}%
\providecommand \BibitemOpen [0]{}%
\providecommand \bibitemStop [0]{}%
\providecommand \bibitemNoStop [0]{.\EOS\space}%
\providecommand \EOS [0]{\spacefactor3000\relax}%
\providecommand \BibitemShut  [1]{\csname bibitem#1\endcsname}%
\let\auto@bib@innerbib\@empty
\bibitem [{\citenamefont {Sugihara}\ and\ \citenamefont {May}(1990)}]{1}%
  \BibitemOpen
  \bibfield  {author} {\bibinfo {author} {\bibfnamefont {G.}~\bibnamefont
  {Sugihara}}\ and\ \bibinfo {author} {\bibfnamefont {R.~M.}\ \bibnamefont
  {May}},\ }\bibfield  {title} {\bibinfo {title} {Nonlinear forecasting as a
  way of distinguishing chaos from measurement error in time series},\
  }\href@noop {} {\bibfield  {journal} {\bibinfo  {journal} {Nature}\ }\textbf
  {\bibinfo {volume} {344}},\ \bibinfo {pages} {734} (\bibinfo {year}
  {1990})}\BibitemShut {NoStop}%
\bibitem [{\citenamefont {Tropp}\ and\ \citenamefont {Gilbert}(2007)}]{2}%
  \BibitemOpen
  \bibfield  {author} {\bibinfo {author} {\bibfnamefont {J.~A.}\ \bibnamefont
  {Tropp}}\ and\ \bibinfo {author} {\bibfnamefont {A.~C.}\ \bibnamefont
  {Gilbert}},\ }\bibfield  {title} {\bibinfo {title} {Signal recovery from
  random measurements via orthogonal matching pursuit},\ }\href@noop {}
  {\bibfield  {journal} {\bibinfo  {journal} {IEEE Transactions on Information
  Theory}\ }\textbf {\bibinfo {volume} {53}},\ \bibinfo {pages} {4655}
  (\bibinfo {year} {2007})}\BibitemShut {NoStop}%
\bibitem [{\citenamefont {Champion}\ \emph {et~al.}(2019)\citenamefont
  {Champion}, \citenamefont {Lusch}, \citenamefont {Kutz},\ and\ \citenamefont
  {Brunton}}]{3}%
  \BibitemOpen
  \bibfield  {author} {\bibinfo {author} {\bibfnamefont {K.}~\bibnamefont
  {Champion}}, \bibinfo {author} {\bibfnamefont {B.}~\bibnamefont {Lusch}},
  \bibinfo {author} {\bibfnamefont {J.~N.}\ \bibnamefont {Kutz}},\ and\
  \bibinfo {author} {\bibfnamefont {S.~L.}\ \bibnamefont {Brunton}},\
  }\bibfield  {title} {\bibinfo {title} {Data-driven discovery of coordinates
  and governing equations},\ }\href@noop {} {\bibfield  {journal} {\bibinfo
  {journal} {Proceedings of the National Academy of Sciences}\ }\textbf
  {\bibinfo {volume} {116}},\ \bibinfo {pages} {22445} (\bibinfo {year}
  {2019})}\BibitemShut {NoStop}%
\bibitem [{\citenamefont {Rudy}\ \emph {et~al.}(2017)\citenamefont {Rudy},
  \citenamefont {Brunton}, \citenamefont {Proctor},\ and\ \citenamefont
  {Kutz}}]{4}%
  \BibitemOpen
  \bibfield  {author} {\bibinfo {author} {\bibfnamefont {S.~H.}\ \bibnamefont
  {Rudy}}, \bibinfo {author} {\bibfnamefont {S.~L.}\ \bibnamefont {Brunton}},
  \bibinfo {author} {\bibfnamefont {J.~L.}\ \bibnamefont {Proctor}},\ and\
  \bibinfo {author} {\bibfnamefont {J.~N.}\ \bibnamefont {Kutz}},\ }\bibfield
  {title} {\bibinfo {title} {Data-driven discovery of partial differential
  equations},\ }\href@noop {} {\bibfield  {journal} {\bibinfo  {journal}
  {Science advances}\ }\textbf {\bibinfo {volume} {3}},\ \bibinfo {pages}
  {e1602614} (\bibinfo {year} {2017})}\BibitemShut {NoStop}%
\bibitem [{\citenamefont {Crutchfield}\ and\ \citenamefont
  {McNamara}(1987)}]{5}%
  \BibitemOpen
  \bibfield  {author} {\bibinfo {author} {\bibfnamefont {J.~P.}\ \bibnamefont
  {Crutchfield}}\ and\ \bibinfo {author} {\bibfnamefont {B.}~\bibnamefont
  {McNamara}},\ }\bibfield  {title} {\bibinfo {title} {Equations of motion from
  a data series},\ }\href@noop {} {\bibfield  {journal} {\bibinfo  {journal}
  {Complex Systems}\ }\textbf {\bibinfo {volume} {1}},\ \bibinfo {pages} {417}
  (\bibinfo {year} {1987})}\BibitemShut {NoStop}%
\bibitem [{\citenamefont {Li}\ \emph {et~al.}(2019)\citenamefont {Li},
  \citenamefont {Kaiser}, \citenamefont {Laima}, \citenamefont {Li},
  \citenamefont {Brunton},\ and\ \citenamefont {Kutz}}]{li2019discovering}%
  \BibitemOpen
  \bibfield  {author} {\bibinfo {author} {\bibfnamefont {S.}~\bibnamefont
  {Li}}, \bibinfo {author} {\bibfnamefont {E.}~\bibnamefont {Kaiser}}, \bibinfo
  {author} {\bibfnamefont {S.}~\bibnamefont {Laima}}, \bibinfo {author}
  {\bibfnamefont {H.}~\bibnamefont {Li}}, \bibinfo {author} {\bibfnamefont
  {S.~L.}\ \bibnamefont {Brunton}},\ and\ \bibinfo {author} {\bibfnamefont
  {J.~N.}\ \bibnamefont {Kutz}},\ }\bibfield  {title} {\bibinfo {title}
  {Discovering time-varying aerodynamics of a prototype bridge by sparse
  identification of nonlinear dynamical systems},\ }\href@noop {} {\bibfield
  {journal} {\bibinfo  {journal} {Physical Review E}\ }\textbf {\bibinfo
  {volume} {100}},\ \bibinfo {pages} {022220} (\bibinfo {year}
  {2019})}\BibitemShut {NoStop}%
\bibitem [{\citenamefont {Reinbold}\ and\ \citenamefont
  {Grigoriev}(2019)}]{reinbold2019data}%
  \BibitemOpen
  \bibfield  {author} {\bibinfo {author} {\bibfnamefont {P.~A.~K.}\
  \bibnamefont {Reinbold}}\ and\ \bibinfo {author} {\bibfnamefont {R.~O.}\
  \bibnamefont {Grigoriev}},\ }\bibfield  {title} {\bibinfo {title}
  {Data-driven discovery of partial differential equation models with latent
  variables},\ }\href@noop {} {\bibfield  {journal} {\bibinfo  {journal}
  {Physical Review E}\ }\textbf {\bibinfo {volume} {100}},\ \bibinfo {pages}
  {022219} (\bibinfo {year} {2019})}\BibitemShut {NoStop}%
\bibitem [{\citenamefont {Maddu}\ \emph {et~al.}(2021)\citenamefont {Maddu},
  \citenamefont {Cheeseman}, \citenamefont {M{\"u}ller},\ and\ \citenamefont
  {Sbalzarini}}]{maddu2021learning}%
  \BibitemOpen
  \bibfield  {author} {\bibinfo {author} {\bibfnamefont {S.}~\bibnamefont
  {Maddu}}, \bibinfo {author} {\bibfnamefont {B.~L.}\ \bibnamefont
  {Cheeseman}}, \bibinfo {author} {\bibfnamefont {C.~L.}\ \bibnamefont
  {M{\"u}ller}},\ and\ \bibinfo {author} {\bibfnamefont {I.~F.}\ \bibnamefont
  {Sbalzarini}},\ }\bibfield  {title} {\bibinfo {title} {Learning physically
  consistent differential equation models from data using group sparsity},\
  }\href@noop {} {\bibfield  {journal} {\bibinfo  {journal} {Physical Review
  E}\ }\textbf {\bibinfo {volume} {103}},\ \bibinfo {pages} {042310} (\bibinfo
  {year} {2021})}\BibitemShut {NoStop}%
\bibitem [{\citenamefont {Somacal}\ \emph {et~al.}(2022)\citenamefont
  {Somacal}, \citenamefont {Barrera}, \citenamefont {Boechi}, \citenamefont
  {Jonckheere}, \citenamefont {Lefieux}, \citenamefont {Picard},\ and\
  \citenamefont {Smucler}}]{somacal2022uncovering}%
  \BibitemOpen
  \bibfield  {author} {\bibinfo {author} {\bibfnamefont {A.}~\bibnamefont
  {Somacal}}, \bibinfo {author} {\bibfnamefont {Y.}~\bibnamefont {Barrera}},
  \bibinfo {author} {\bibfnamefont {L.}~\bibnamefont {Boechi}}, \bibinfo
  {author} {\bibfnamefont {M.}~\bibnamefont {Jonckheere}}, \bibinfo {author}
  {\bibfnamefont {V.}~\bibnamefont {Lefieux}}, \bibinfo {author} {\bibfnamefont
  {D.}~\bibnamefont {Picard}},\ and\ \bibinfo {author} {\bibfnamefont
  {E.}~\bibnamefont {Smucler}},\ }\bibfield  {title} {\bibinfo {title}
  {Uncovering differential equations from data with hidden variables},\
  }\href@noop {} {\bibfield  {journal} {\bibinfo  {journal} {Physical Review
  E}\ }\textbf {\bibinfo {volume} {105}},\ \bibinfo {pages} {054209} (\bibinfo
  {year} {2022})}\BibitemShut {NoStop}%
\bibitem [{\citenamefont {Shea}\ \emph {et~al.}(2021)\citenamefont {Shea},
  \citenamefont {Brunton},\ and\ \citenamefont {Kutz}}]{shea2021sindy}%
  \BibitemOpen
  \bibfield  {author} {\bibinfo {author} {\bibfnamefont {D.~E.}\ \bibnamefont
  {Shea}}, \bibinfo {author} {\bibfnamefont {S.~L.}\ \bibnamefont {Brunton}},\
  and\ \bibinfo {author} {\bibfnamefont {J.~N.}\ \bibnamefont {Kutz}},\
  }\bibfield  {title} {\bibinfo {title} {Sindy-bvp: Sparse identification of
  nonlinear dynamics for boundary value problems},\ }\href@noop {} {\bibfield
  {journal} {\bibinfo  {journal} {Physical Review Research}\ }\textbf {\bibinfo
  {volume} {3}},\ \bibinfo {pages} {023255} (\bibinfo {year}
  {2021})}\BibitemShut {NoStop}%
\bibitem [{\citenamefont {Xu}\ and\ \citenamefont
  {Zhang}(2021)}]{xu2021robust}%
  \BibitemOpen
  \bibfield  {author} {\bibinfo {author} {\bibfnamefont {H.}~\bibnamefont
  {Xu}}\ and\ \bibinfo {author} {\bibfnamefont {D.}~\bibnamefont {Zhang}},\
  }\bibfield  {title} {\bibinfo {title} {Robust discovery of partial
  differential equations in complex situations},\ }\href@noop {} {\bibfield
  {journal} {\bibinfo  {journal} {Physical Review Research}\ }\textbf {\bibinfo
  {volume} {3}},\ \bibinfo {pages} {033270} (\bibinfo {year}
  {2021})}\BibitemShut {NoStop}%
\bibitem [{\citenamefont {Chen}\ \emph {et~al.}(2022)\citenamefont {Chen},
  \citenamefont {Luo}, \citenamefont {Liu}, \citenamefont {Xu},\ and\
  \citenamefont {Zhang}}]{chen2022symbolic}%
  \BibitemOpen
  \bibfield  {author} {\bibinfo {author} {\bibfnamefont {Y.}~\bibnamefont
  {Chen}}, \bibinfo {author} {\bibfnamefont {Y.}~\bibnamefont {Luo}}, \bibinfo
  {author} {\bibfnamefont {Q.}~\bibnamefont {Liu}}, \bibinfo {author}
  {\bibfnamefont {H.}~\bibnamefont {Xu}},\ and\ \bibinfo {author}
  {\bibfnamefont {D.}~\bibnamefont {Zhang}},\ }\bibfield  {title} {\bibinfo
  {title} {Symbolic genetic algorithm for discovering open-form partial
  differential equations (sga-pde)},\ }\href@noop {} {\bibfield  {journal}
  {\bibinfo  {journal} {Physical Review Research}\ }\textbf {\bibinfo {volume}
  {4}},\ \bibinfo {pages} {023174} (\bibinfo {year} {2022})}\BibitemShut
  {NoStop}%
\bibitem [{\citenamefont {Ye}\ \emph {et~al.}(2015)\citenamefont {Ye},
  \citenamefont {Beamish}, \citenamefont {Glaser}, \citenamefont {Grant},
  \citenamefont {Hsieh}, \citenamefont {Richards}, \citenamefont {Schnute},\
  and\ \citenamefont {Sugihara}}]{6}%
  \BibitemOpen
  \bibfield  {author} {\bibinfo {author} {\bibfnamefont {H.}~\bibnamefont
  {Ye}}, \bibinfo {author} {\bibfnamefont {R.~J.}\ \bibnamefont {Beamish}},
  \bibinfo {author} {\bibfnamefont {S.~M.}\ \bibnamefont {Glaser}}, \bibinfo
  {author} {\bibfnamefont {S.~C.}\ \bibnamefont {Grant}}, \bibinfo {author}
  {\bibfnamefont {C.-h.}\ \bibnamefont {Hsieh}}, \bibinfo {author}
  {\bibfnamefont {L.~J.}\ \bibnamefont {Richards}}, \bibinfo {author}
  {\bibfnamefont {J.~T.}\ \bibnamefont {Schnute}},\ and\ \bibinfo {author}
  {\bibfnamefont {G.}~\bibnamefont {Sugihara}},\ }\bibfield  {title} {\bibinfo
  {title} {Equation-free mechanistic ecosystem forecasting using empirical
  dynamic modeling},\ }\href@noop {} {\bibfield  {journal} {\bibinfo  {journal}
  {Proceedings of the National Academy of Sciences}\ }\textbf {\bibinfo
  {volume} {112}},\ \bibinfo {pages} {E1569} (\bibinfo {year}
  {2015})}\BibitemShut {NoStop}%
\bibitem [{\citenamefont {Kevrekidis}\ \emph {et~al.}(2003)\citenamefont
  {Kevrekidis}, \citenamefont {Gear}, \citenamefont {Hyman}, \citenamefont
  {Kevrekidis}, \citenamefont {Runborg}, \citenamefont {Theodoropoulos} \emph
  {et~al.}}]{7}%
  \BibitemOpen
  \bibfield  {author} {\bibinfo {author} {\bibfnamefont {I.~G.}\ \bibnamefont
  {Kevrekidis}}, \bibinfo {author} {\bibfnamefont {C.~W.}\ \bibnamefont
  {Gear}}, \bibinfo {author} {\bibfnamefont {J.~M.}\ \bibnamefont {Hyman}},
  \bibinfo {author} {\bibfnamefont {P.~G.}\ \bibnamefont {Kevrekidis}},
  \bibinfo {author} {\bibfnamefont {O.}~\bibnamefont {Runborg}}, \bibinfo
  {author} {\bibfnamefont {C.}~\bibnamefont {Theodoropoulos}}, \emph {et~al.},\
  }\bibfield  {title} {\bibinfo {title} {Equation-free, coarse-grained
  multiscale computation: enabling microscopic simulators to perform
  system-level analysis},\ }\href@noop {} {\bibfield  {journal} {\bibinfo
  {journal} {Commun. Math. Sci}\ }\textbf {\bibinfo {volume} {1}},\ \bibinfo
  {pages} {715} (\bibinfo {year} {2003})}\BibitemShut {NoStop}%
\bibitem [{\citenamefont {Proctor}\ \emph {et~al.}(2014)\citenamefont
  {Proctor}, \citenamefont {Brunton}, \citenamefont {Brunton},\ and\
  \citenamefont {Kutz}}]{8}%
  \BibitemOpen
  \bibfield  {author} {\bibinfo {author} {\bibfnamefont {J.~L.}\ \bibnamefont
  {Proctor}}, \bibinfo {author} {\bibfnamefont {S.~L.}\ \bibnamefont
  {Brunton}}, \bibinfo {author} {\bibfnamefont {B.~W.}\ \bibnamefont
  {Brunton}},\ and\ \bibinfo {author} {\bibfnamefont {J.}~\bibnamefont
  {Kutz}},\ }\bibfield  {title} {\bibinfo {title} {Exploiting sparsity and
  equation-free architectures in complex systems},\ }\href@noop {} {\bibfield
  {journal} {\bibinfo  {journal} {The European Physical Journal Special
  Topics}\ }\textbf {\bibinfo {volume} {223}},\ \bibinfo {pages} {2665}
  (\bibinfo {year} {2014})}\BibitemShut {NoStop}%
\bibitem [{\citenamefont {Roberts}(2014)}]{9}%
  \BibitemOpen
  \bibfield  {author} {\bibinfo {author} {\bibfnamefont {A.~J.}\ \bibnamefont
  {Roberts}},\ }\href@noop {} {\emph {\bibinfo {title} {Model emergent dynamics
  in complex systems}}},\ Vol.~\bibinfo {volume} {20}\ (\bibinfo  {publisher}
  {SIAM},\ \bibinfo {year} {2014})\BibitemShut {NoStop}%
\bibitem [{\citenamefont {Majda}\ \emph {et~al.}(2009)\citenamefont {Majda},
  \citenamefont {Franzke},\ and\ \citenamefont {Crommelin}}]{10}%
  \BibitemOpen
  \bibfield  {author} {\bibinfo {author} {\bibfnamefont {A.~J.}\ \bibnamefont
  {Majda}}, \bibinfo {author} {\bibfnamefont {C.}~\bibnamefont {Franzke}},\
  and\ \bibinfo {author} {\bibfnamefont {D.}~\bibnamefont {Crommelin}},\
  }\bibfield  {title} {\bibinfo {title} {Normal forms for reduced stochastic
  climate models},\ }\href@noop {} {\bibfield  {journal} {\bibinfo  {journal}
  {Proceedings of the National Academy of Sciences}\ }\textbf {\bibinfo
  {volume} {106}},\ \bibinfo {pages} {3649} (\bibinfo {year}
  {2009})}\BibitemShut {NoStop}%
\bibitem [{\citenamefont {Giannakis}\ and\ \citenamefont {Majda}(2012)}]{11}%
  \BibitemOpen
  \bibfield  {author} {\bibinfo {author} {\bibfnamefont {D.}~\bibnamefont
  {Giannakis}}\ and\ \bibinfo {author} {\bibfnamefont {A.~J.}\ \bibnamefont
  {Majda}},\ }\bibfield  {title} {\bibinfo {title} {Nonlinear laplacian
  spectral analysis for time series with intermittency and low-frequency
  variability},\ }\href@noop {} {\bibfield  {journal} {\bibinfo  {journal}
  {Proceedings of the National Academy of Sciences}\ }\textbf {\bibinfo
  {volume} {109}},\ \bibinfo {pages} {2222} (\bibinfo {year}
  {2012})}\BibitemShut {NoStop}%
\bibitem [{\citenamefont {Daniels}\ and\ \citenamefont {Nemenman}(2015)}]{12}%
  \BibitemOpen
  \bibfield  {author} {\bibinfo {author} {\bibfnamefont {B.~C.}\ \bibnamefont
  {Daniels}}\ and\ \bibinfo {author} {\bibfnamefont {I.}~\bibnamefont
  {Nemenman}},\ }\bibfield  {title} {\bibinfo {title} {Automated adaptive
  inference of phenomenological dynamical models},\ }\href@noop {} {\bibfield
  {journal} {\bibinfo  {journal} {Nature Communications}\ }\textbf {\bibinfo
  {volume} {6}},\ \bibinfo {pages} {1} (\bibinfo {year} {2015})}\BibitemShut
  {NoStop}%
\bibitem [{\citenamefont {Voss}\ \emph {et~al.}(1999)\citenamefont {Voss},
  \citenamefont {Kolodner}, \citenamefont {Abel},\ and\ \citenamefont
  {Kurths}}]{13}%
  \BibitemOpen
  \bibfield  {author} {\bibinfo {author} {\bibfnamefont {H.~U.}\ \bibnamefont
  {Voss}}, \bibinfo {author} {\bibfnamefont {P.}~\bibnamefont {Kolodner}},
  \bibinfo {author} {\bibfnamefont {M.}~\bibnamefont {Abel}},\ and\ \bibinfo
  {author} {\bibfnamefont {J.}~\bibnamefont {Kurths}},\ }\bibfield  {title}
  {\bibinfo {title} {Amplitude equations from spatiotemporal binary-fluid
  convection data},\ }\href@noop {} {\bibfield  {journal} {\bibinfo  {journal}
  {Physical Review Letters}\ }\textbf {\bibinfo {volume} {83}},\ \bibinfo
  {pages} {3422} (\bibinfo {year} {1999})}\BibitemShut {NoStop}%
\bibitem [{\citenamefont {Tibshirani}(1996)}]{14}%
  \BibitemOpen
  \bibfield  {author} {\bibinfo {author} {\bibfnamefont {R.}~\bibnamefont
  {Tibshirani}},\ }\bibfield  {title} {\bibinfo {title} {Regression shrinkage
  and selection via the lasso},\ }\href@noop {} {\bibfield  {journal} {\bibinfo
   {journal} {Journal of the Royal Statistical Society. Series B
  (Methodological)}\ }\textbf {\bibinfo {volume} {58}},\ \bibinfo {pages} {267}
  (\bibinfo {year} {1996})}\BibitemShut {NoStop}%
\bibitem [{\citenamefont {Mangan}\ \emph {et~al.}(2016)\citenamefont {Mangan},
  \citenamefont {Brunton}, \citenamefont {Proctor},\ and\ \citenamefont
  {Kutz}}]{15}%
  \BibitemOpen
  \bibfield  {author} {\bibinfo {author} {\bibfnamefont {N.~M.}\ \bibnamefont
  {Mangan}}, \bibinfo {author} {\bibfnamefont {S.~L.}\ \bibnamefont {Brunton}},
  \bibinfo {author} {\bibfnamefont {J.~L.}\ \bibnamefont {Proctor}},\ and\
  \bibinfo {author} {\bibfnamefont {J.~N.}\ \bibnamefont {Kutz}},\ }\bibfield
  {title} {\bibinfo {title} {Inferring biological networks by sparse
  identification of nonlinear dynamics},\ }\href@noop {} {\bibfield  {journal}
  {\bibinfo  {journal} {IEEE Transactions on Molecular, Biological and
  Multi-Scale Communications}\ }\textbf {\bibinfo {volume} {2}},\ \bibinfo
  {pages} {52} (\bibinfo {year} {2016})}\BibitemShut {NoStop}%
\bibitem [{\citenamefont {Brunton}\ and\ \citenamefont {Kutz}(2022)}]{16}%
  \BibitemOpen
  \bibfield  {author} {\bibinfo {author} {\bibfnamefont {S.~L.}\ \bibnamefont
  {Brunton}}\ and\ \bibinfo {author} {\bibfnamefont {J.~N.}\ \bibnamefont
  {Kutz}},\ }\href@noop {} {\emph {\bibinfo {title} {Data-driven science and
  engineering: Machine learning, dynamical systems, and control}}}\ (\bibinfo
  {publisher} {Cambridge University Press},\ \bibinfo {year}
  {2022})\BibitemShut {NoStop}%
\bibitem [{\citenamefont {Lusch}\ \emph {et~al.}(2018)\citenamefont {Lusch},
  \citenamefont {Kutz},\ and\ \citenamefont {Brunton}}]{18}%
  \BibitemOpen
  \bibfield  {author} {\bibinfo {author} {\bibfnamefont {B.}~\bibnamefont
  {Lusch}}, \bibinfo {author} {\bibfnamefont {J.~N.}\ \bibnamefont {Kutz}},\
  and\ \bibinfo {author} {\bibfnamefont {S.~L.}\ \bibnamefont {Brunton}},\
  }\bibfield  {title} {\bibinfo {title} {Deep learning for universal linear
  embeddings of nonlinear dynamics},\ }\href@noop {} {\bibfield  {journal}
  {\bibinfo  {journal} {Nature communications}\ }\textbf {\bibinfo {volume}
  {9}},\ \bibinfo {pages} {1} (\bibinfo {year} {2018})}\BibitemShut {NoStop}%
\bibitem [{\citenamefont {Brunton}\ \emph {et~al.}(2016)\citenamefont
  {Brunton}, \citenamefont {Proctor},\ and\ \citenamefont {Kutz}}]{20}%
  \BibitemOpen
  \bibfield  {author} {\bibinfo {author} {\bibfnamefont {S.~L.}\ \bibnamefont
  {Brunton}}, \bibinfo {author} {\bibfnamefont {J.~L.}\ \bibnamefont
  {Proctor}},\ and\ \bibinfo {author} {\bibfnamefont {J.~N.}\ \bibnamefont
  {Kutz}},\ }\bibfield  {title} {\bibinfo {title} {Discovering governing
  equations from data by sparse identification of nonlinear dynamical
  systems},\ }\href@noop {} {\bibfield  {journal} {\bibinfo  {journal}
  {Proceedings of the National Academy of Sciences}\ }\textbf {\bibinfo
  {volume} {113}},\ \bibinfo {pages} {3932} (\bibinfo {year}
  {2016})}\BibitemShut {NoStop}%
\bibitem [{\citenamefont {Long}\ \emph {et~al.}(2018)\citenamefont {Long},
  \citenamefont {Lu}, \citenamefont {Ma},\ and\ \citenamefont {Dong}}]{19}%
  \BibitemOpen
  \bibfield  {author} {\bibinfo {author} {\bibfnamefont {Z.}~\bibnamefont
  {Long}}, \bibinfo {author} {\bibfnamefont {Y.}~\bibnamefont {Lu}}, \bibinfo
  {author} {\bibfnamefont {X.}~\bibnamefont {Ma}},\ and\ \bibinfo {author}
  {\bibfnamefont {B.}~\bibnamefont {Dong}},\ }\bibfield  {title} {\bibinfo
  {title} {Pde-net: Learning pdes from data},\ }in\ \href@noop {} {\emph
  {\bibinfo {booktitle} {International Conference on Machine Learning}}}\
  (\bibinfo {organization} {PMLR},\ \bibinfo {year} {2018})\ pp.\ \bibinfo
  {pages} {3208--3216}\BibitemShut {NoStop}%
\bibitem [{\citenamefont {Hallac}\ \emph {et~al.}(2017)\citenamefont {Hallac},
  \citenamefont {Park}, \citenamefont {Boyd},\ and\ \citenamefont
  {Leskovec}}]{21}%
  \BibitemOpen
  \bibfield  {author} {\bibinfo {author} {\bibfnamefont {D.}~\bibnamefont
  {Hallac}}, \bibinfo {author} {\bibfnamefont {Y.}~\bibnamefont {Park}},
  \bibinfo {author} {\bibfnamefont {S.}~\bibnamefont {Boyd}},\ and\ \bibinfo
  {author} {\bibfnamefont {J.}~\bibnamefont {Leskovec}},\ }\bibfield  {title}
  {\bibinfo {title} {Network inference via the time-varying graphical lasso},\
  }in\ \href@noop {} {\emph {\bibinfo {booktitle} {Proceedings of the 23rd ACM
  SIGKDD International Conference on Knowledge Discovery and Data Mining}}}\
  (\bibinfo {year} {2017})\ pp.\ \bibinfo {pages} {205--213}\BibitemShut
  {NoStop}%
\bibitem [{\citenamefont {Ranjan}(2014)}]{30}%
  \BibitemOpen
  \bibfield  {author} {\bibinfo {author} {\bibfnamefont {R.}~\bibnamefont
  {Ranjan}},\ }\bibfield  {title} {\bibinfo {title} {Modeling and simulation in
  performance optimization of big data processing frameworks},\ }\href@noop {}
  {\bibfield  {journal} {\bibinfo  {journal} {IEEE Cloud Computing}\ }\textbf
  {\bibinfo {volume} {1}},\ \bibinfo {pages} {14} (\bibinfo {year}
  {2014})}\BibitemShut {NoStop}%
\bibitem [{\citenamefont {Namaki}\ \emph {et~al.}(2011)\citenamefont {Namaki},
  \citenamefont {Shirazi}, \citenamefont {Raei},\ and\ \citenamefont
  {Jafari}}]{25}%
  \BibitemOpen
  \bibfield  {author} {\bibinfo {author} {\bibfnamefont {A.}~\bibnamefont
  {Namaki}}, \bibinfo {author} {\bibfnamefont {A.~H.}\ \bibnamefont {Shirazi}},
  \bibinfo {author} {\bibfnamefont {R.}~\bibnamefont {Raei}},\ and\ \bibinfo
  {author} {\bibfnamefont {G.}~\bibnamefont {Jafari}},\ }\bibfield  {title}
  {\bibinfo {title} {Network analysis of a financial market based on genuine
  correlation and threshold method},\ }\href@noop {} {\bibfield  {journal}
  {\bibinfo  {journal} {Physica A: Statistical Mechanics and its Applications}\
  }\textbf {\bibinfo {volume} {390}},\ \bibinfo {pages} {3835} (\bibinfo {year}
  {2011})}\BibitemShut {NoStop}%
\bibitem [{\citenamefont {Ahmed}\ and\ \citenamefont {Xing}(2009)}]{26}%
  \BibitemOpen
  \bibfield  {author} {\bibinfo {author} {\bibfnamefont {A.}~\bibnamefont
  {Ahmed}}\ and\ \bibinfo {author} {\bibfnamefont {E.~P.}\ \bibnamefont
  {Xing}},\ }\bibfield  {title} {\bibinfo {title} {Recovering time-varying
  networks of dependencies in social and biological studies},\ }\href@noop {}
  {\bibfield  {journal} {\bibinfo  {journal} {Proceedings of the National
  Academy of Sciences}\ }\textbf {\bibinfo {volume} {106}},\ \bibinfo {pages}
  {11878} (\bibinfo {year} {2009})}\BibitemShut {NoStop}%
\bibitem [{\citenamefont {Myers}\ and\ \citenamefont {Leskovec}(2010)}]{27}%
  \BibitemOpen
  \bibfield  {author} {\bibinfo {author} {\bibfnamefont {S.}~\bibnamefont
  {Myers}}\ and\ \bibinfo {author} {\bibfnamefont {J.}~\bibnamefont
  {Leskovec}},\ }\bibfield  {title} {\bibinfo {title} {On the convexity of
  latent social network inference},\ }\href@noop {} {\bibfield  {journal}
  {\bibinfo  {journal} {Advances in Neural Information Processing systems}\
  }\textbf {\bibinfo {volume} {23}} (\bibinfo {year} {2010})}\BibitemShut
  {NoStop}%
\bibitem [{\citenamefont {Monti}\ \emph {et~al.}(2014)\citenamefont {Monti},
  \citenamefont {Hellyer}, \citenamefont {Sharp}, \citenamefont {Leech},
  \citenamefont {Anagnostopoulos},\ and\ \citenamefont {Montana}}]{28}%
  \BibitemOpen
  \bibfield  {author} {\bibinfo {author} {\bibfnamefont {R.~P.}\ \bibnamefont
  {Monti}}, \bibinfo {author} {\bibfnamefont {P.}~\bibnamefont {Hellyer}},
  \bibinfo {author} {\bibfnamefont {D.}~\bibnamefont {Sharp}}, \bibinfo
  {author} {\bibfnamefont {R.}~\bibnamefont {Leech}}, \bibinfo {author}
  {\bibfnamefont {C.}~\bibnamefont {Anagnostopoulos}},\ and\ \bibinfo {author}
  {\bibfnamefont {G.}~\bibnamefont {Montana}},\ }\bibfield  {title} {\bibinfo
  {title} {Estimating time-varying brain connectivity networks from functional
  mri time series},\ }\href@noop {} {\bibfield  {journal} {\bibinfo  {journal}
  {NeuroImage}\ }\textbf {\bibinfo {volume} {103}},\ \bibinfo {pages} {427}
  (\bibinfo {year} {2014})}\BibitemShut {NoStop}%
\bibitem [{\citenamefont {Cai}\ \emph {et~al.}(2022)\citenamefont {Cai},
  \citenamefont {Wang}, \citenamefont {Joos},\ and\ \citenamefont
  {Kamwa}}]{cai2022online}%
  \BibitemOpen
  \bibfield  {author} {\bibinfo {author} {\bibfnamefont {Y.}~\bibnamefont
  {Cai}}, \bibinfo {author} {\bibfnamefont {X.}~\bibnamefont {Wang}}, \bibinfo
  {author} {\bibfnamefont {G.}~\bibnamefont {Joos}},\ and\ \bibinfo {author}
  {\bibfnamefont {I.}~\bibnamefont {Kamwa}},\ }\bibfield  {title} {\bibinfo
  {title} {An online data-driven method to locate forced oscillation sources
  from power plants based on sparse identification of nonlinear dynamics
  (sindy)},\ }\href@noop {} {\bibfield  {journal} {\bibinfo  {journal} {IEEE
  Transactions on Power Systems}\ } (\bibinfo {year} {2022})}\BibitemShut
  {NoStop}%
\bibitem [{\citenamefont {Conti}\ \emph {et~al.}(2023)\citenamefont {Conti},
  \citenamefont {Gobat}, \citenamefont {Fresca}, \citenamefont {Manzoni},\ and\
  \citenamefont {Frangi}}]{conti2023reduced}%
  \BibitemOpen
  \bibfield  {author} {\bibinfo {author} {\bibfnamefont {P.}~\bibnamefont
  {Conti}}, \bibinfo {author} {\bibfnamefont {G.}~\bibnamefont {Gobat}},
  \bibinfo {author} {\bibfnamefont {S.}~\bibnamefont {Fresca}}, \bibinfo
  {author} {\bibfnamefont {A.}~\bibnamefont {Manzoni}},\ and\ \bibinfo {author}
  {\bibfnamefont {A.}~\bibnamefont {Frangi}},\ }\bibfield  {title} {\bibinfo
  {title} {Reduced order modeling of parametrized systems through autoencoders
  and sindy approach: continuation of periodic solutions},\ }\href@noop {}
  {\bibfield  {journal} {\bibinfo  {journal} {Computer Methods in Applied
  Mechanics and Engineering}\ }\textbf {\bibinfo {volume} {411}},\ \bibinfo
  {pages} {116072} (\bibinfo {year} {2023})}\BibitemShut {NoStop}%
\bibitem [{\citenamefont {Messenger}\ \emph {et~al.}(2022)\citenamefont
  {Messenger}, \citenamefont {Dall’Anese},\ and\ \citenamefont
  {Bortz}}]{messenger2022online}%
  \BibitemOpen
  \bibfield  {author} {\bibinfo {author} {\bibfnamefont {D.~A.}\ \bibnamefont
  {Messenger}}, \bibinfo {author} {\bibfnamefont {E.}~\bibnamefont
  {Dall’Anese}},\ and\ \bibinfo {author} {\bibfnamefont {D.}~\bibnamefont
  {Bortz}},\ }\bibfield  {title} {\bibinfo {title} {Online weak-form sparse
  identification of partial differential equations},\ }in\ \href@noop {} {\emph
  {\bibinfo {booktitle} {Mathematical and Scientific Machine Learning}}}\
  (\bibinfo {organization} {PMLR},\ \bibinfo {year} {2022})\ pp.\ \bibinfo
  {pages} {241--256}\BibitemShut {NoStop}%
\bibitem [{\citenamefont {Schaeffer}\ \emph {et~al.}(2013)\citenamefont
  {Schaeffer}, \citenamefont {Caflisch}, \citenamefont {Hauck},\ and\
  \citenamefont {Osher}}]{32}%
  \BibitemOpen
  \bibfield  {author} {\bibinfo {author} {\bibfnamefont {H.}~\bibnamefont
  {Schaeffer}}, \bibinfo {author} {\bibfnamefont {R.}~\bibnamefont {Caflisch}},
  \bibinfo {author} {\bibfnamefont {C.~D.}\ \bibnamefont {Hauck}},\ and\
  \bibinfo {author} {\bibfnamefont {S.}~\bibnamefont {Osher}},\ }\bibfield
  {title} {\bibinfo {title} {Sparse dynamics for partial differential
  equations},\ }\href@noop {} {\bibfield  {journal} {\bibinfo  {journal}
  {Proceedings of the National Academy of Sciences}\ }\textbf {\bibinfo
  {volume} {110}},\ \bibinfo {pages} {6634} (\bibinfo {year}
  {2013})}\BibitemShut {NoStop}%
\bibitem [{\citenamefont {McMahan}(2011)}]{34}%
  \BibitemOpen
  \bibfield  {author} {\bibinfo {author} {\bibfnamefont {B.}~\bibnamefont
  {McMahan}},\ }\bibfield  {title} {\bibinfo {title}
  {Follow-the-regularized-leader and mirror descent: Equivalence theorems and
  l1 regularization},\ }in\ \href@noop {} {\emph {\bibinfo {booktitle}
  {Proceedings of the Fourteenth International Conference on Artificial
  Intelligence and Statistics}}}\ (\bibinfo {year} {2011})\ pp.\ \bibinfo
  {pages} {525--533}\BibitemShut {NoStop}%
\bibitem [{\citenamefont {McMahan}\ \emph {et~al.}(2013)\citenamefont
  {McMahan}, \citenamefont {Holt}, \citenamefont {Sculley}, \citenamefont
  {Young}, \citenamefont {Ebner}, \citenamefont {Grady}, \citenamefont {Nie},
  \citenamefont {Phillips}, \citenamefont {Davydov}, \citenamefont {Golovin}
  \emph {et~al.}}]{35}%
  \BibitemOpen
  \bibfield  {author} {\bibinfo {author} {\bibfnamefont {H.~B.}\ \bibnamefont
  {McMahan}}, \bibinfo {author} {\bibfnamefont {G.}~\bibnamefont {Holt}},
  \bibinfo {author} {\bibfnamefont {D.}~\bibnamefont {Sculley}}, \bibinfo
  {author} {\bibfnamefont {M.}~\bibnamefont {Young}}, \bibinfo {author}
  {\bibfnamefont {D.}~\bibnamefont {Ebner}}, \bibinfo {author} {\bibfnamefont
  {J.}~\bibnamefont {Grady}}, \bibinfo {author} {\bibfnamefont
  {L.}~\bibnamefont {Nie}}, \bibinfo {author} {\bibfnamefont {T.}~\bibnamefont
  {Phillips}}, \bibinfo {author} {\bibfnamefont {E.}~\bibnamefont {Davydov}},
  \bibinfo {author} {\bibfnamefont {D.}~\bibnamefont {Golovin}}, \emph
  {et~al.},\ }\bibfield  {title} {\bibinfo {title} {Ad click prediction: a view
  from the trenches},\ }in\ \href@noop {} {\emph {\bibinfo {booktitle}
  {Proceedings of the 19th ACM SIGKDD International Conference on Knowledge
  Discovery and Data Mining}}}\ (\bibinfo {year} {2013})\ pp.\ \bibinfo {pages}
  {1222--1230}\BibitemShut {NoStop}%
\bibitem [{\citenamefont {Wu}\ \emph {et~al.}(2022)\citenamefont {Wu},
  \citenamefont {Hao}, \citenamefont {Liu}, \citenamefont {Liu},\ and\
  \citenamefont {Shen}}]{36}%
  \BibitemOpen
  \bibfield  {author} {\bibinfo {author} {\bibfnamefont {K.}~\bibnamefont
  {Wu}}, \bibinfo {author} {\bibfnamefont {X.}~\bibnamefont {Hao}}, \bibinfo
  {author} {\bibfnamefont {J.}~\bibnamefont {Liu}}, \bibinfo {author}
  {\bibfnamefont {P.}~\bibnamefont {Liu}},\ and\ \bibinfo {author}
  {\bibfnamefont {F.}~\bibnamefont {Shen}},\ }\bibfield  {title} {\bibinfo
  {title} {Online reconstruction of complex networks from streaming data},\
  }\href {https://doi.org/10.1109/TCYB.2020.3027642} {\bibfield  {journal}
  {\bibinfo  {journal} {IEEE Transactions on Cybernetics}\ }\textbf {\bibinfo
  {volume} {52}},\ \bibinfo {pages} {5136} (\bibinfo {year}
  {2022})}\BibitemShut {NoStop}%
\bibitem [{\citenamefont {Chen}\ \emph {et~al.}(2001)\citenamefont {Chen},
  \citenamefont {Donoho},\ and\ \citenamefont {Saunders}}]{39}%
  \BibitemOpen
  \bibfield  {author} {\bibinfo {author} {\bibfnamefont {S.~S.}\ \bibnamefont
  {Chen}}, \bibinfo {author} {\bibfnamefont {D.~L.}\ \bibnamefont {Donoho}},\
  and\ \bibinfo {author} {\bibfnamefont {M.~A.}\ \bibnamefont {Saunders}},\
  }\bibfield  {title} {\bibinfo {title} {Atomic decomposition by basis
  pursuit},\ }\href@noop {} {\bibfield  {journal} {\bibinfo  {journal} {SIAM
  Review}\ }\textbf {\bibinfo {volume} {43}},\ \bibinfo {pages} {129} (\bibinfo
  {year} {2001})}\BibitemShut {NoStop}%
\bibitem [{Note1()}]{Note1}%
  \BibitemOpen
  \bibinfo {note} {Available code:
  https://github.com/xiaoyuans/O-SINDy.}\BibitemShut {Stop}%
\bibitem [{\citenamefont {Budi{\v{s}}i{\'c}}\ \emph {et~al.}(2012)\citenamefont
  {Budi{\v{s}}i{\'c}}, \citenamefont {Mohr},\ and\ \citenamefont
  {Mezi{\'c}}}]{42}%
  \BibitemOpen
  \bibfield  {author} {\bibinfo {author} {\bibfnamefont {M.}~\bibnamefont
  {Budi{\v{s}}i{\'c}}}, \bibinfo {author} {\bibfnamefont {R.}~\bibnamefont
  {Mohr}},\ and\ \bibinfo {author} {\bibfnamefont {I.}~\bibnamefont
  {Mezi{\'c}}},\ }\bibfield  {title} {\bibinfo {title} {Applied koopmanism},\
  }\href@noop {} {\bibfield  {journal} {\bibinfo  {journal} {Chaos: An
  Interdisciplinary Journal of Nonlinear Science}\ }\textbf {\bibinfo {volume}
  {22}},\ \bibinfo {pages} {047510} (\bibinfo {year} {2012})}\BibitemShut
  {NoStop}%
\bibitem [{\citenamefont {Zou}\ and\ \citenamefont {Hastie}(2005)}]{43}%
  \BibitemOpen
  \bibfield  {author} {\bibinfo {author} {\bibfnamefont {H.}~\bibnamefont
  {Zou}}\ and\ \bibinfo {author} {\bibfnamefont {T.}~\bibnamefont {Hastie}},\
  }\bibfield  {title} {\bibinfo {title} {Regularization and variable selection
  via the elastic net},\ }\href@noop {} {\bibfield  {journal} {\bibinfo
  {journal} {Journal of the royal statistical society: series B (statistical
  methodology)}\ }\textbf {\bibinfo {volume} {67}},\ \bibinfo {pages} {301}
  (\bibinfo {year} {2005})}\BibitemShut {NoStop}%
\bibitem [{\citenamefont {Zhang}(2008)}]{44}%
  \BibitemOpen
  \bibfield  {author} {\bibinfo {author} {\bibfnamefont {T.}~\bibnamefont
  {Zhang}},\ }\bibfield  {title} {\bibinfo {title} {Adaptive forward-backward
  greedy algorithm for sparse learning with linear models},\ }\href@noop {}
  {\bibfield  {journal} {\bibinfo  {journal} {Advances in neural information
  processing systems}\ }\textbf {\bibinfo {volume} {21}},\ \bibinfo {pages}
  {1921–1928} (\bibinfo {year} {2008})}\BibitemShut {NoStop}%
\bibitem [{\citenamefont {Grenfell}\ \emph {et~al.}(1994)\citenamefont
  {Grenfell}, \citenamefont {Kleckzkowski}, \citenamefont {Ellner},\ and\
  \citenamefont {Bolker}}]{grenfell1994measles}%
  \BibitemOpen
  \bibfield  {author} {\bibinfo {author} {\bibfnamefont {B.~T.}\ \bibnamefont
  {Grenfell}}, \bibinfo {author} {\bibfnamefont {A.}~\bibnamefont
  {Kleckzkowski}}, \bibinfo {author} {\bibfnamefont {S.}~\bibnamefont
  {Ellner}},\ and\ \bibinfo {author} {\bibfnamefont {B.}~\bibnamefont
  {Bolker}},\ }\bibfield  {title} {\bibinfo {title} {Measles as a case study in
  nonlinear forecasting and chaos},\ }\href@noop {} {\bibfield  {journal}
  {\bibinfo  {journal} {Philosophical Transactions of the Royal Society of
  London. Series A: Physical and Engineering Sciences}\ }\textbf {\bibinfo
  {volume} {348}},\ \bibinfo {pages} {515} (\bibinfo {year}
  {1994})}\BibitemShut {NoStop}%
\bibitem [{\citenamefont {Mangan}\ \emph {et~al.}(2019)\citenamefont {Mangan},
  \citenamefont {Askham}, \citenamefont {Brunton}, \citenamefont {Kutz},\ and\
  \citenamefont {Proctor}}]{mangan2019model}%
  \BibitemOpen
  \bibfield  {author} {\bibinfo {author} {\bibfnamefont {N.~M.}\ \bibnamefont
  {Mangan}}, \bibinfo {author} {\bibfnamefont {T.}~\bibnamefont {Askham}},
  \bibinfo {author} {\bibfnamefont {S.~L.}\ \bibnamefont {Brunton}}, \bibinfo
  {author} {\bibfnamefont {J.~N.}\ \bibnamefont {Kutz}},\ and\ \bibinfo
  {author} {\bibfnamefont {J.~L.}\ \bibnamefont {Proctor}},\ }\bibfield
  {title} {\bibinfo {title} {Model selection for hybrid dynamical systems via
  sparse regression},\ }\href@noop {} {\bibfield  {journal} {\bibinfo
  {journal} {Proceedings of the Royal Society A}\ }\textbf {\bibinfo {volume}
  {475}},\ \bibinfo {pages} {20180534} (\bibinfo {year} {2019})}\BibitemShut
  {NoStop}%
\bibitem [{\citenamefont {Harris}\ \emph {et~al.}(2021)\citenamefont {Harris},
  \citenamefont {Aravkin}, \citenamefont {Rao},\ and\ \citenamefont
  {Brunton}}]{harris2021time}%
  \BibitemOpen
  \bibfield  {author} {\bibinfo {author} {\bibfnamefont {K.~D.}\ \bibnamefont
  {Harris}}, \bibinfo {author} {\bibfnamefont {A.}~\bibnamefont {Aravkin}},
  \bibinfo {author} {\bibfnamefont {R.}~\bibnamefont {Rao}},\ and\ \bibinfo
  {author} {\bibfnamefont {B.~W.}\ \bibnamefont {Brunton}},\ }\bibfield
  {title} {\bibinfo {title} {Time-varying autoregression with low-rank
  tensors},\ }\href@noop {} {\bibfield  {journal} {\bibinfo  {journal} {SIAM
  Journal on Applied Dynamical Systems}\ }\textbf {\bibinfo {volume} {20}},\
  \bibinfo {pages} {2335} (\bibinfo {year} {2021})}\BibitemShut {NoStop}%
\bibitem [{\citenamefont {Stephens}\ \emph {et~al.}(2008)\citenamefont
  {Stephens}, \citenamefont {Johnson-Kerner}, \citenamefont {Bialek},\ and\
  \citenamefont {Ryu}}]{stephens2008dimensionality}%
  \BibitemOpen
  \bibfield  {author} {\bibinfo {author} {\bibfnamefont {G.~J.}\ \bibnamefont
  {Stephens}}, \bibinfo {author} {\bibfnamefont {B.}~\bibnamefont
  {Johnson-Kerner}}, \bibinfo {author} {\bibfnamefont {W.}~\bibnamefont
  {Bialek}},\ and\ \bibinfo {author} {\bibfnamefont {W.~S.}\ \bibnamefont
  {Ryu}},\ }\bibfield  {title} {\bibinfo {title} {Dimensionality and dynamics
  in the behavior of c. elegans},\ }\href@noop {} {\bibfield  {journal}
  {\bibinfo  {journal} {PLoS computational biology}\ }\textbf {\bibinfo
  {volume} {4}},\ \bibinfo {pages} {e1000028} (\bibinfo {year}
  {2008})}\BibitemShut {NoStop}%
\bibitem [{\citenamefont {Broekmans}\ \emph {et~al.}(2016)\citenamefont
  {Broekmans}, \citenamefont {Rodgers}, \citenamefont {Ryu},\ and\
  \citenamefont {Stephens}}]{broekmans2016resolving}%
  \BibitemOpen
  \bibfield  {author} {\bibinfo {author} {\bibfnamefont {O.~D.}\ \bibnamefont
  {Broekmans}}, \bibinfo {author} {\bibfnamefont {J.~B.}\ \bibnamefont
  {Rodgers}}, \bibinfo {author} {\bibfnamefont {W.~S.}\ \bibnamefont {Ryu}},\
  and\ \bibinfo {author} {\bibfnamefont {G.~J.}\ \bibnamefont {Stephens}},\
  }\bibfield  {title} {\bibinfo {title} {Resolving coiled shapes reveals new
  reorientation behaviors in c. elegans},\ }\href@noop {} {\bibfield  {journal}
  {\bibinfo  {journal} {Elife}\ }\textbf {\bibinfo {volume} {5}},\ \bibinfo
  {pages} {e17227} (\bibinfo {year} {2016})}\BibitemShut {NoStop}%
\bibitem [{\citenamefont {Costa}\ \emph {et~al.}(2019)\citenamefont {Costa},
  \citenamefont {Ahamed},\ and\ \citenamefont {Stephens}}]{costa2019adaptive}%
  \BibitemOpen
  \bibfield  {author} {\bibinfo {author} {\bibfnamefont {A.~C.}\ \bibnamefont
  {Costa}}, \bibinfo {author} {\bibfnamefont {T.}~\bibnamefont {Ahamed}},\ and\
  \bibinfo {author} {\bibfnamefont {G.~J.}\ \bibnamefont {Stephens}},\
  }\bibfield  {title} {\bibinfo {title} {Adaptive, locally linear models of
  complex dynamics},\ }\href@noop {} {\bibfield  {journal} {\bibinfo  {journal}
  {Proceedings of the National Academy of Sciences}\ }\textbf {\bibinfo
  {volume} {116}},\ \bibinfo {pages} {1501} (\bibinfo {year}
  {2019})}\BibitemShut {NoStop}%
\bibitem [{\citenamefont {LeVeque}(2007)}]{37}%
  \BibitemOpen
  \bibfield  {author} {\bibinfo {author} {\bibfnamefont {R.~J.}\ \bibnamefont
  {LeVeque}},\ }\href@noop {} {\emph {\bibinfo {title} {Finite difference
  methods for ordinary and partial differential equations: steady-state and
  time-dependent problems}}}\ (\bibinfo  {publisher} {SIAM},\ \bibinfo {year}
  {2007})\BibitemShut {NoStop}%
\bibitem [{\citenamefont {Bruno}\ and\ \citenamefont {Hoch}(2012)}]{38}%
  \BibitemOpen
  \bibfield  {author} {\bibinfo {author} {\bibfnamefont {O.}~\bibnamefont
  {Bruno}}\ and\ \bibinfo {author} {\bibfnamefont {D.}~\bibnamefont {Hoch}},\
  }\bibfield  {title} {\bibinfo {title} {Numerical differentiation of
  approximated functions with limited order-of-accuracy deterioration},\
  }\href@noop {} {\bibfield  {journal} {\bibinfo  {journal} {SIAM Journal on
  Numerical Analysis}\ }\textbf {\bibinfo {volume} {50}},\ \bibinfo {pages}
  {1581} (\bibinfo {year} {2012})}\BibitemShut {NoStop}%
\bibitem [{\citenamefont {Gavish}\ and\ \citenamefont {Donoho}(2014)}]{40}%
  \BibitemOpen
  \bibfield  {author} {\bibinfo {author} {\bibfnamefont {M.}~\bibnamefont
  {Gavish}}\ and\ \bibinfo {author} {\bibfnamefont {D.~L.}\ \bibnamefont
  {Donoho}},\ }\bibfield  {title} {\bibinfo {title} {The optimal hard threshold
  for singular values is $4/sqrt{3}$},\ }\href@noop {} {\bibfield  {journal}
  {\bibinfo  {journal} {IEEE Transactions on Information Theory}\ }\textbf
  {\bibinfo {volume} {60}},\ \bibinfo {pages} {5040} (\bibinfo {year}
  {2014})}\BibitemShut {NoStop}%
\bibitem [{\citenamefont {Walters}(1987)}]{22}%
  \BibitemOpen
  \bibfield  {author} {\bibinfo {author} {\bibfnamefont {C.~J.}\ \bibnamefont
  {Walters}},\ }\bibfield  {title} {\bibinfo {title} {Nonstationarity of
  production relationships in exploited populations},\ }\href@noop {}
  {\bibfield  {journal} {\bibinfo  {journal} {Canadian Journal of Fisheries and
  Aquatic Sciences}\ }\textbf {\bibinfo {volume} {44}},\ \bibinfo {pages}
  {s156} (\bibinfo {year} {1987})}\BibitemShut {NoStop}%
\bibitem [{\citenamefont {Sugihara}\ \emph {et~al.}(2012)\citenamefont
  {Sugihara}, \citenamefont {May}, \citenamefont {Ye}, \citenamefont {Hsieh},
  \citenamefont {Deyle}, \citenamefont {Fogarty},\ and\ \citenamefont
  {Munch}}]{23}%
  \BibitemOpen
  \bibfield  {author} {\bibinfo {author} {\bibfnamefont {G.}~\bibnamefont
  {Sugihara}}, \bibinfo {author} {\bibfnamefont {R.}~\bibnamefont {May}},
  \bibinfo {author} {\bibfnamefont {H.}~\bibnamefont {Ye}}, \bibinfo {author}
  {\bibfnamefont {C.-h.}\ \bibnamefont {Hsieh}}, \bibinfo {author}
  {\bibfnamefont {E.}~\bibnamefont {Deyle}}, \bibinfo {author} {\bibfnamefont
  {M.}~\bibnamefont {Fogarty}},\ and\ \bibinfo {author} {\bibfnamefont
  {S.}~\bibnamefont {Munch}},\ }\bibfield  {title} {\bibinfo {title} {Detecting
  causality in complex ecosystems},\ }\href@noop {} {\bibfield  {journal}
  {\bibinfo  {journal} {Science}\ }\textbf {\bibinfo {volume} {338}},\ \bibinfo
  {pages} {496} (\bibinfo {year} {2012})}\BibitemShut {NoStop}%
\bibitem [{\citenamefont {Zhu}\ and\ \citenamefont {Xu}(2005)}]{24}%
  \BibitemOpen
  \bibfield  {author} {\bibinfo {author} {\bibfnamefont {X.}~\bibnamefont
  {Zhu}}\ and\ \bibinfo {author} {\bibfnamefont {J.}~\bibnamefont {Xu}},\
  }\bibfield  {title} {\bibinfo {title} {Estimation of time varying parameters
  in nonlinear systems by using dynamic optimization},\ }in\ \href
  {https://doi.org/10.1109/IECON.2005.1568871} {\emph {\bibinfo {booktitle}
  {31st Annual Conference of IEEE Industrial Electronics Society, 2005. IECON
  2005.}}}\ (\bibinfo {year} {2005})\ pp.\ \bibinfo {pages} {5
  pp.--}\BibitemShut {NoStop}%
\bibitem [{\citenamefont {Kolar}\ \emph {et~al.}(2010)\citenamefont {Kolar},
  \citenamefont {Song}, \citenamefont {Ahmed},\ and\ \citenamefont
  {Xing}}]{29}%
  \BibitemOpen
  \bibfield  {author} {\bibinfo {author} {\bibfnamefont {M.}~\bibnamefont
  {Kolar}}, \bibinfo {author} {\bibfnamefont {L.}~\bibnamefont {Song}},
  \bibinfo {author} {\bibfnamefont {A.}~\bibnamefont {Ahmed}},\ and\ \bibinfo
  {author} {\bibfnamefont {E.~P.}\ \bibnamefont {Xing}},\ }\bibfield  {title}
  {\bibinfo {title} {Estimating time-varying networks},\ }\href@noop {}
  {\bibfield  {journal} {\bibinfo  {journal} {The Annals of Applied
  Statistics}\ ,\ \bibinfo {pages} {94}} (\bibinfo {year} {2010})}\BibitemShut
  {NoStop}%
\bibitem [{\citenamefont {Deyle}\ \emph {et~al.}(2013)\citenamefont {Deyle},
  \citenamefont {Fogarty}, \citenamefont {Hsieh}, \citenamefont {Kaufman},
  \citenamefont {MacCall}, \citenamefont {Munch}, \citenamefont {Perretti},
  \citenamefont {Ye},\ and\ \citenamefont {Sugihara}}]{31}%
  \BibitemOpen
  \bibfield  {author} {\bibinfo {author} {\bibfnamefont {E.~R.}\ \bibnamefont
  {Deyle}}, \bibinfo {author} {\bibfnamefont {M.}~\bibnamefont {Fogarty}},
  \bibinfo {author} {\bibfnamefont {C.-h.}\ \bibnamefont {Hsieh}}, \bibinfo
  {author} {\bibfnamefont {L.}~\bibnamefont {Kaufman}}, \bibinfo {author}
  {\bibfnamefont {A.~D.}\ \bibnamefont {MacCall}}, \bibinfo {author}
  {\bibfnamefont {S.~B.}\ \bibnamefont {Munch}}, \bibinfo {author}
  {\bibfnamefont {C.~T.}\ \bibnamefont {Perretti}}, \bibinfo {author}
  {\bibfnamefont {H.}~\bibnamefont {Ye}},\ and\ \bibinfo {author}
  {\bibfnamefont {G.}~\bibnamefont {Sugihara}},\ }\bibfield  {title} {\bibinfo
  {title} {Predicting climate effects on pacific sardine},\ }\href@noop {}
  {\bibfield  {journal} {\bibinfo  {journal} {Proceedings of the National
  Academy of Sciences}\ }\textbf {\bibinfo {volume} {110}},\ \bibinfo {pages}
  {6430} (\bibinfo {year} {2013})}\BibitemShut {NoStop}%
\bibitem [{\citenamefont {Cand{\`e}s}\ \emph {et~al.}(2006)\citenamefont
  {Cand{\`e}s}, \citenamefont {Romberg},\ and\ \citenamefont {Tao}}]{33}%
  \BibitemOpen
  \bibfield  {author} {\bibinfo {author} {\bibfnamefont {E.~J.}\ \bibnamefont
  {Cand{\`e}s}}, \bibinfo {author} {\bibfnamefont {J.}~\bibnamefont
  {Romberg}},\ and\ \bibinfo {author} {\bibfnamefont {T.}~\bibnamefont {Tao}},\
  }\bibfield  {title} {\bibinfo {title} {Robust uncertainty principles: Exact
  signal reconstruction from highly incomplete frequency information},\
  }\href@noop {} {\bibfield  {journal} {\bibinfo  {journal} {IEEE Transactions
  on Information Theory}\ }\textbf {\bibinfo {volume} {52}},\ \bibinfo {pages}
  {489} (\bibinfo {year} {2006})}\BibitemShut {NoStop}%
\bibitem [{\citenamefont {Ruelle}\ and\ \citenamefont {Takens}(1971)}]{41}%
  \BibitemOpen
  \bibfield  {author} {\bibinfo {author} {\bibfnamefont {D.}~\bibnamefont
  {Ruelle}}\ and\ \bibinfo {author} {\bibfnamefont {F.}~\bibnamefont
  {Takens}},\ }\bibfield  {title} {\bibinfo {title} {On the nature of
  turbulence},\ }\href@noop {} {\bibfield  {journal} {\bibinfo  {journal}
  {Communications in Mathematical Physics}\ }\textbf {\bibinfo {volume} {20}},\
  \bibinfo {pages} {167} (\bibinfo {year} {1971})}\BibitemShut {NoStop}%
\bibitem [{\citenamefont {Bramburger}\ \emph {et~al.}(2020)\citenamefont
  {Bramburger}, \citenamefont {Dylewsky},\ and\ \citenamefont
  {Kutz}}]{bramburger2020sparse}%
  \BibitemOpen
  \bibfield  {author} {\bibinfo {author} {\bibfnamefont {J.~J.}\ \bibnamefont
  {Bramburger}}, \bibinfo {author} {\bibfnamefont {D.}~\bibnamefont
  {Dylewsky}},\ and\ \bibinfo {author} {\bibfnamefont {J.~N.}\ \bibnamefont
  {Kutz}},\ }\bibfield  {title} {\bibinfo {title} {Sparse identification of
  slow timescale dynamics},\ }\href@noop {} {\bibfield  {journal} {\bibinfo
  {journal} {Physical Review E}\ }\textbf {\bibinfo {volume} {102}},\ \bibinfo
  {pages} {022204} (\bibinfo {year} {2020})}\BibitemShut {NoStop}%
\bibitem [{\citenamefont {Dylewsky}\ \emph {et~al.}(2022)\citenamefont
  {Dylewsky}, \citenamefont {Kaiser}, \citenamefont {Brunton},\ and\
  \citenamefont {Kutz}}]{dylewsky2022principal}%
  \BibitemOpen
  \bibfield  {author} {\bibinfo {author} {\bibfnamefont {D.}~\bibnamefont
  {Dylewsky}}, \bibinfo {author} {\bibfnamefont {E.}~\bibnamefont {Kaiser}},
  \bibinfo {author} {\bibfnamefont {S.~L.}\ \bibnamefont {Brunton}},\ and\
  \bibinfo {author} {\bibfnamefont {J.~N.}\ \bibnamefont {Kutz}},\ }\bibfield
  {title} {\bibinfo {title} {Principal component trajectories for modeling
  spectrally continuous dynamics as forced linear systems},\ }\href@noop {}
  {\bibfield  {journal} {\bibinfo  {journal} {Physical Review E}\ }\textbf
  {\bibinfo {volume} {105}},\ \bibinfo {pages} {015312} (\bibinfo {year}
  {2022})}\BibitemShut {NoStop}%
\bibitem [{\citenamefont {Tran}\ \emph {et~al.}(2023)\citenamefont {Tran},
  \citenamefont {Tran}, \citenamefont {Nguyen},\ and\ \citenamefont
  {Pemen}}]{tran2023sparse}%
  \BibitemOpen
  \bibfield  {author} {\bibinfo {author} {\bibfnamefont {M.-Q.}\ \bibnamefont
  {Tran}}, \bibinfo {author} {\bibfnamefont {T.~T.}\ \bibnamefont {Tran}},
  \bibinfo {author} {\bibfnamefont {P.~H.}\ \bibnamefont {Nguyen}},\ and\
  \bibinfo {author} {\bibfnamefont {G.}~\bibnamefont {Pemen}},\ }\bibfield
  {title} {\bibinfo {title} {Sparse identification for model predictive control
  to support long-term voltage stability},\ }\href@noop {} {\bibfield
  {journal} {\bibinfo  {journal} {IET Generation, Transmission \&
  Distribution}\ }\textbf {\bibinfo {volume} {17}},\ \bibinfo {pages} {39}
  (\bibinfo {year} {2023})}\BibitemShut {NoStop}%
\end{thebibliography}

%

\end{document}